\documentclass[a4paper,11pt]{article}
 \usepackage{jheppub} 
\usepackage{graphicx}
 \usepackage{amssymb}
\usepackage{dcolumn}
\usepackage{bm}
\usepackage{color}
\usepackage{multirow}
 \newcommand{\ud}{\mathrm{d}}
\newcommand{\met}{{/\!\!\! E_T}} 
\newcommand{\mpt}{{/\!\!\!\! \vec{P}_T}} 
\allowdisplaybreaks

 \newcommand{\lsim}{{\;\raise0.3ex\hbox{$<$\kern-0.75em\raise-1.1ex\hbox{$\sim$}}\;}}
\newcommand{\gsim}{{\;\raise0.3ex\hbox{$>$\kern-0.75em\raise-1.1ex\hbox{$\sim$}}\;}}
\newcommand{\beq}{\begin{equation}}
\newcommand{\eeq}{\end{equation}}
\newcommand{\bea}{\begin{eqnarray}}
\newcommand{\eea}{\end{eqnarray}}
\mathchardef\minus="002D

\title{\boldmath Re-interpreting the Oxbridge stransverse mass variable $M_{T2}$ in general cases}

\author[a]{Rakhi Mahbubani,}
\author[b]{Konstantin T.~Matchev,}  
\author[a]{Myeonghun Park,\note{Corresponding author: {\tt Myeonghun.Park@cern.ch}}}

\affiliation[a]{CERN Physics Department, Theory Division, CH-1211 Geneva 23, Switzerland.}
\affiliation[b]{Physics Department, University of Florida, Gainesville, FL 32611, USA.}

\abstract{We extend the range of possible applications of $M_{T2}$ type analyses 
to decay chains with multiple invisible particles, as well as to asymmetric event topologies 
with different parent and/or different children particles. We advocate two possible approaches.
In the first, we introduce suitably defined $3+1$-dimensional analogues of the $M_{T2}$
variable, which take into account all relevant on-shell kinematic constraints in a given event topology.
The second approach utilizes the conventional $M_{T2}$ variable, but its kinematic endpoint
is suitably reinterpreted on a case by case basis, depending on the specific event topology at hand. 
We provide the general prescription for this reinterpretation, including the formulas relating the 
measured $M_{T2}$ endpoint (as a function of the test masses of all the invisible particles) 
to the underlying physical mass spectrum.
We also provide analytical formulas for the shape of the differential distribution of the 
doubly projected $M_{T2_\perp}$ variable for the ten possible event topologies with 
one visible particle and up to two invisible particles per decay chain.
We illustrate our results with the example of leptonic chargino decays
$\tilde\chi^+\to \ell^+ \nu \tilde \chi^0$ in supersymmetry.}

\date{December 3, 2012}
\preprint{CERN-PH-TH/2012-333}  

\begin{document} 
\maketitle
\flushbottom

\section{Introduction}
\label{sec:introduction}

\begin{figure}[t]
\begin{center}
\includegraphics[width=10cm]{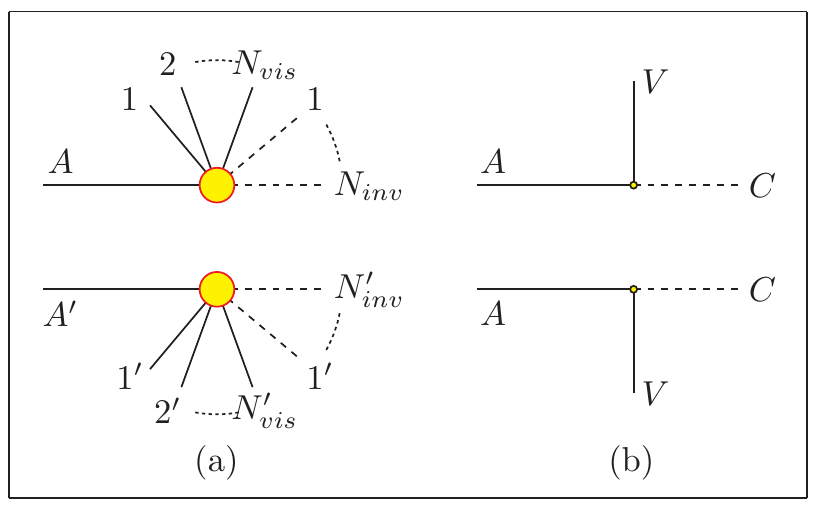}
\end{center}
\caption{\label{fig:example} (a) The generic event topology for the pair production of two parent particles $A$ and $A'$.
Particle $A$ ($A'$) decays to a set of $N_{vis}$ ($N'_{vis}$) particles which are visible in the detector and a set of
$N_{inv}$ ($N'_{inv}$) invisible particles. The invisible particles are not necessarily limited to SM neutrinos, 
and can have arbitrary masses.
(b) The simplified event topology typically used in $M_{T2}$ studies. 
Each $V$ is either a single SM particle, or an effective composite visible particle 
which is constructed from the corresponding visible particles in figure (a). 
The decay chain for each parent $A$ terminates in a single invisible particle $C$.}
\end{figure}

The Large Hadron Collider (LHC) at CERN is continuing its quest for new physics Beyond the Standard Model (BSM).
Among the multitude of possible BSM scenarios, models with neutral and stable WIMPs (weakly interacting massive 
particles) are of particular interest. First, such models are greatly motivated by the dark matter problem, as WIMPs are 
suitable dark matter candidates. Second, the initial BSM searches at the LHC have already
placed stringent limits on heavy resonances which decay visibly and can be fully reconstructed. 
In contrast, the limits on parent particles which decay {\em semi-invisibly} 
(to a collection of visible SM particles and one or more WIMPs) are much weaker.
First, the background issue in these analyses is more complicated: since the parent cannot be fully reconstructed, 
the search is not a mere ``bump hunt", where the background can be simply subtracted from the side-bands. 
Second, the symmetry which protects the lifetime of the WIMP dark matter candidate 
typically requires that the new particles are multiply produced, leading to lower production cross-sections
(as opposed to single production). In the simplest and most popular models, the new symmetry is a ${\cal Z}_2$ parity,
which implies that the new particles are pair produced, as illustrated in Fig.\,\ref{fig:example}.

Fig.\,\ref{fig:example}(a) depicts the generic event topology in a typical SUSY-like missing energy event.
Two parent particles $A$ and $A'$ are produced, and each one decays to a certain number 
of visible particles (denoted by solid lines), and a certain number of invisible particles
(denoted by dashed lines). At this point, the diagram in Fig.\,\ref{fig:example}(a) is meant to represent 
the most general case: for example, the number of visible particles ($N_{vis}$ and $N'_{vis}$), 
as well as the number of invisible particles ($N_{inv}$ and $N'_{inv}$) in each parent decay chain 
is completely arbitrary. Furthermore, the invisible particles are not necessarily limited to SM neutrinos, 
but may include (several species of) WIMPs with different masses.
Most importantly, no assumption has been made regarding the actual decay topology of the parents, 
and the yellow-shaded circles are simply placeholders for the actual $1\to N_{vis}+N_{inv}$ and $1\to N'_{vis}+N'_{inv}$
Feynman diagrams responsible for the parent decays in the figure.

The kinematic analysis of events of this type is rather challenging. There is a lot of missing (or a priori unknown) 
information: the number of invisible particles $N_{inv}$ and $N'_{inv}$, the momenta (and masses) of the invisible 
particles, the total invariant mass and longitudinal momentum of the parent system $(A,A')$,
and the event topologies hiding behind the yellow-shaded circles in Fig.\,\ref{fig:example}(a). 
These problems have inspired a lot of previous work in the literature on new methods for measuring 
the masses of the parents and the invisible daughters under these circumstances (for a recent review, see \cite{Barr:2010zj}).
Among the different options existing in the literature, the {\em invariant mass} variables appear to be 
both useful and theoretically motivated\,\cite{Barr:2011xt}. Being Lorentz-invariant, they are insensitive to the
a priori unknown longitudinal boost of the $(A, A')$ system. A special subset of invariant mass variables which
also shares this property, is given by the {\em transverse} invariant mass variables, among which the  
Oxbridge stransverse mass $M_{T2}$\,\cite{Lester:1999tx,Barr:2003rg} is a well-known example directly
applicable to Fig.\,\ref{fig:example}(a).
It has been featured in analyses performed by the Tevatron and LHC experimental collaborations CDF, D0, CMS and ATLAS\,\cite{Aaltonen:2009rm,daCosta:2011qk,Chatrchyan:2012jx,CMStop,Abazov:2012}.  

In the original proposal\,\cite{Lester:1999tx}, the ``Oxbridge" $M_{T2}$ variable was applied to the case of
direct slepton production, corresponding to the simple event topology shown in Fig.\,\ref{fig:example}(b), which
is a special case of the more general Fig.\,\ref{fig:example}(a). In Fig.\,\ref{fig:example}(b), two identical 
parents ($A$) are produced, and each one subsequently decays via a two-body decay to a single visible SM particle 
$V$ and a single invisible child particle $C$. In spite of its simplicity, the event topology of Fig.\,\ref{fig:example}(b)
covers a number of interesting physics cases, e.g. slepton production\,\cite{Lester:1999tx,Barr:2005dz}, 
squark production\,\cite{Randall:2008rw,Nojiri:2011qn}, chargino production\,\cite {Matchev:2009fh,Belanger:2011ny}, etc. 
At the same time, the $M_{T2}$ concept is very powerful, and can be usefully applied to situations that
are more general than the simple example in Fig.\,\ref{fig:example}(b). 

So let us first review the different directions in which one could generalize Fig.\,\ref{fig:example}(b). 
Some of the following options (a-b) have already been considered in the literature, and we only mention them here for completeness.
The remaining possibilities (c-f), however, have attracted significantly less attention in the literature, and will be the 
main focus of this paper.
\begin{itemize}
\item[(a)] In Fig.\,\ref{fig:example}(b) there is only one visible particle in each decay chain. In the language of
Fig.\,\ref{fig:example}(a), this implies the assumption $N_{vis}=N'_{vis}=1$. At the same type, a typical BSM model
like supersymmetry exhibits much longer decay chains, with several visible particles on each side of the event.
This case can be easily handled with the conventional $M_{T2}$ approach --- one just needs to 
think of each $V$ as a collection of visible particles with some net four-momentum $P^\mu$, which is measured 
in the detector. Early work along these lines\,\cite{Cho:2007qv,Cho:2007dh,Barr:2007hy} led to the discovery of the
$M_{T2}$ ``kink" in the endpoint $M_{T2}^{max}$ when considered as a function of the unknown test\footnote{Following the
notation of \cite{Burns:2008va}, test input masses for invisible particles will be denoted by a tilde.} 
mass $\tilde M_C$ for the child particle:
\beq
\left(\frac{d M_{T2}^{max}(\tilde M_C)}{d\tilde M_C}\right)_{\tilde M_C=M_C(1-\epsilon)} \ne 
\left(\frac{d M_{T2}^{max}(\tilde M_C)}{d\tilde M_C}\right)_{\tilde M_C=M_C(1+\epsilon)}.
\label{eq:kink}
\eeq
The kink in eq.\,(\ref{eq:kink}) is an interesting and unique property of the $M_{T2}$ variable, allowing to measure
simultaneously (at least as a matter of principle) the true mass $M_C$ of the invisible dark matter candidate $C$ at the end of the decay chain {\em and} the true mass $M_A$ of the parent particle initiating the decay chain.
What makes the appearance of the kink possible is the fact that the invariant mass $M_V$ of the visible collection
of particles $V$ is not constant, but varies from event to event: $M_V^2=P^\mu P_\mu \ne const$.
\item[(b)] Another possible generalization is to consider that the parent particles $A$ are produced {\em inclusively},
either in association with jets from initial state radiation, or in the decays of other, even heavier, new particles.
In either case, one again finds an $M_{T2}$ kink in eq.\,(\ref{eq:kink}) at the true mass of the
daughter particle $C$\,\cite{Barr:2007hy,Burns:2008va}. The kink persists even if the decay 
chain is extremely short ($N_{vis}=N'_{vis}=1$) and $M_V = const$. This is because the origin of the 
kink is now different --- it is due to the net transverse momentum of the parent system $(A, A)$.
\item[(c)] Another limiting assumption in Fig.\,\ref{fig:example}(b) is that the two missing particles $C$ 
at the end of each decay chain are the same (or at least have a common mass $M_C$). This assumption, 
however, can be easily relaxed --- one simply needs to allow for two independent mass inputs $\tilde M_{C_1}$
and $\tilde M_{C_2}$ in the calculation of $M_{T2}$\,\cite{Konar:2009qr}. The resulting ``asymmetric"
$M_{T2}$ variable inherits all the desired properties of the conventional $M_{T2}$.
In particular, the ``kink'' in eq.\,(\ref{eq:kink}) in the function $M_{T2}^{max}(\tilde M_C)$ is generalized to a kinky
``crease'' in the surface defined by the function $M_{T2}^{max}(\tilde M_{C_1},\tilde M_{C_2})$\,\cite{Konar:2009qr,Barr:2009jv}.
While this all sounds very straightforward, one should keep in mind that the original public codes\,\cite{MT2library,Cheng:2008hk} for calculating the $M_{T2}$ variable cannot be used in this case, since the
assumption $\tilde M_{C_1}=\tilde M_{C_2}$ is already hardwired\footnote{However, see 
\cite{Bai:2012gs} for an update to \cite{Cheng:2008hk}.}.
\item[(d)] The other assumption in Fig.\,\ref{fig:example}(b) is that the two
parent particles are the same (or at least have a common mass $M_A$).
In principle, this can also be handled rather easily. If the parents are different, but the children are the same,
then one of the decay chains must have additional visible particles which are not present on the other side.
Then, one possibility is to try to identify the extra particles and remove them from consideration, thus reducing the
effective event topology back to Fig.\,\ref{fig:example}(b)\,\cite{Nojiri:2008vq}. An alternative strategy is to 
consider the full event, but allow for different parent masses $\tilde M_{A_1}$ and $\tilde M_{A_2}$
and then use $M_{T2}^{max}$ to construct the function $\tilde M_{C}(\tilde M_{A_1},\tilde M_{A_2})$\,\cite{Barr:2009jv}. 
\item[(e)] Another assumption of Fig.\,\ref{fig:example}(b) is that there is a {\em single} invisible particle 
in each decay chain, i.e.~that $N_{inv}=N'_{inv}=1$. Again, there is no compelling reason for this
restriction in light of generic BSM models. 
First, a ${\cal Z}_N$ parity restricts the number of dark matter
candidates in each decay chain only modulo $N$. Second, a ${\cal Z}_3$ parity can lead to $N_{inv}=2$
as easily as $N_{inv}=1$\,\cite{Agashe:2010tu}. Finally, and most importantly, we already know that 
SM neutrinos exist and behave like invisible particles at colliders. The decay chains in Fig.\,\ref{fig:example}
can easily\footnote{The standard example is the chargino decay in supersymmetry 
$\tilde \chi^\pm_1\to \ell^\pm \nu \tilde\chi^0_1$, which gives {\em two} invisible particles --- 
a neutrino $\nu$ and a neutralino $\tilde\chi^0_1$.} contain SM neutrino particles, 
which would contribute to the total invisible particle count $N_{inv}$\,\cite{Giudice:2011ib}.
Under those circumstances, one needs to generalize the $M_{T2}$ variable to account for the 
extra invisible particles, and define corresponding variables $M_{T3}$, $M_{T4}$, etc.\,\cite{Barr:2003rg},
where the numerical subscript indicates the number of invisible particles $N_{inv}$.
In doing so, however, one faces the following fundamental problem. Recall that the most useful property 
of the original $M_{T2}$ variable was that its endpoint $M_{T2}^{max}$ {\em equals} the true parent mass $M_A$ 
when the test child mass $\tilde M_C$ coincides with the true child mass $M_C$:
\beq
M_{T2}^{max} (\tilde M_{C_i}=M_{C_i}) = M_A.
\label{MT2maxsaturated}
\eeq
However, as we shall explicitly see below in Section~\ref{sec:effective}, this property depends crucially
on the assumption that there aren't any additional invisible particles in the game. In reality, whenever 
the collection of particles $V$ in Fig.\,\ref{fig:example}(b) contains other invisible particles
$\chi$ {\em in addition to} the invisible child particle $C$, and there are intermediate on-shell resonances 
in the cascade decay chain, the bound of eq.\,(\ref{MT2maxsaturated}) 
is not necessarily saturated, and one can only write
\beq
M_{T2}^{max} (M_\chi+M_C) \le M_A.
\label{MT2maxNOTsaturated}
\eeq  
The inequality applies even when the true masses $M_\chi$ and $M_C$ are being used.
The same holds for the new ``generalized'' variables $M_{T3}$, $M_{T4}$, etc.~from \cite{Barr:2003rg}.
In fact, these generalized variables are simply related to the asymmetric $M_{T2}$ variable\,\cite{Konar:2009qr}
as follows\footnote{Here 
and below our notation is that the additional invisible particle $\chi_i$ appears in the same decay chain as the 
invisible child particle $C_i$. See Fig.\,\ref{fig:diagrams} for explicit examples.}
\bea
M_{T3}(\tilde M_{C_1}, \tilde M_{\chi_1}; \tilde M_{C_2}) &=& M_{T2}(\tilde M_{C_1}+ \tilde M_{\chi_1},  \tilde M_{C_2}),   \\
M_{T4}(\tilde M_{C_1}, \tilde M_{\chi_1}; \tilde M_{C_2}, \tilde M_{\chi_2}) 
&=& M_{T2}(\tilde M_{C_1}+ \tilde M_{\chi_1},  \tilde M_{C_2}+ \tilde M_{\chi_2}),  
\eea
and so on. Therefore, the endpoints of $M_{T3}$, $M_{T4}$, etc.~are also {\em not} guaranteed 
to provide a saturated bound like eq.\,(\ref{MT2maxsaturated}) and instead
the best one can do with them is to put a lower limit on $M_A$ as in eq.\,(\ref{MT2maxNOTsaturated}). 
One of the main goals of this paper, therefore, will be to address the problem of
multiple invisible particles and propose how to recover {\em saturated} bounds of the type in eq.\,(\ref{MT2maxsaturated}). 
\item[(f)] The final issue is inherently related to the previous point: once we allow for multiple invisible particles ($N_{inv}>1$)
in the decay chains, we must also address the question of the correct event topology.
In other words, we must ask the question, which Feynman diagram is hiding behind the 
yellow-shaded placeholder in Fig.\,\ref{fig:example}(a). Most studies in the literature
already assume that the correct event topology is known and rarely discuss what happens 
when this assumption is incorrect\,\cite{Blanke:2010cm,Rajaraman:2010hy,Bai:2010hd,Baringer:2011nh,Choi:2011ys}.
Consider, for example, the simplest possible case of one visible particle on each side ($N_{vis}=N'_{vis}=1$) and
then let us allow up to two invisible particles in a given decay chain ($N_{inv}\le 2$ and $N'_{inv}\le2$).
Even for this simple case, there are 10 possible decay topologies which are explicitly shown in Fig.\,\ref{fig:diagrams}(a-j). 
Furthermore, since one cannot be absolutely sure that the two visible particles originated from 
opposite decay chains, in principle one should also contemplate event topologies with 
$N_{vis}=2$ and $N'_{vis}=0$, and two such examples are shown in Fig.\,\ref{fig:diagrams}(k-l).
\end{itemize}

\begin{figure}[t] 
\begin{center}
\includegraphics[width=14cm]{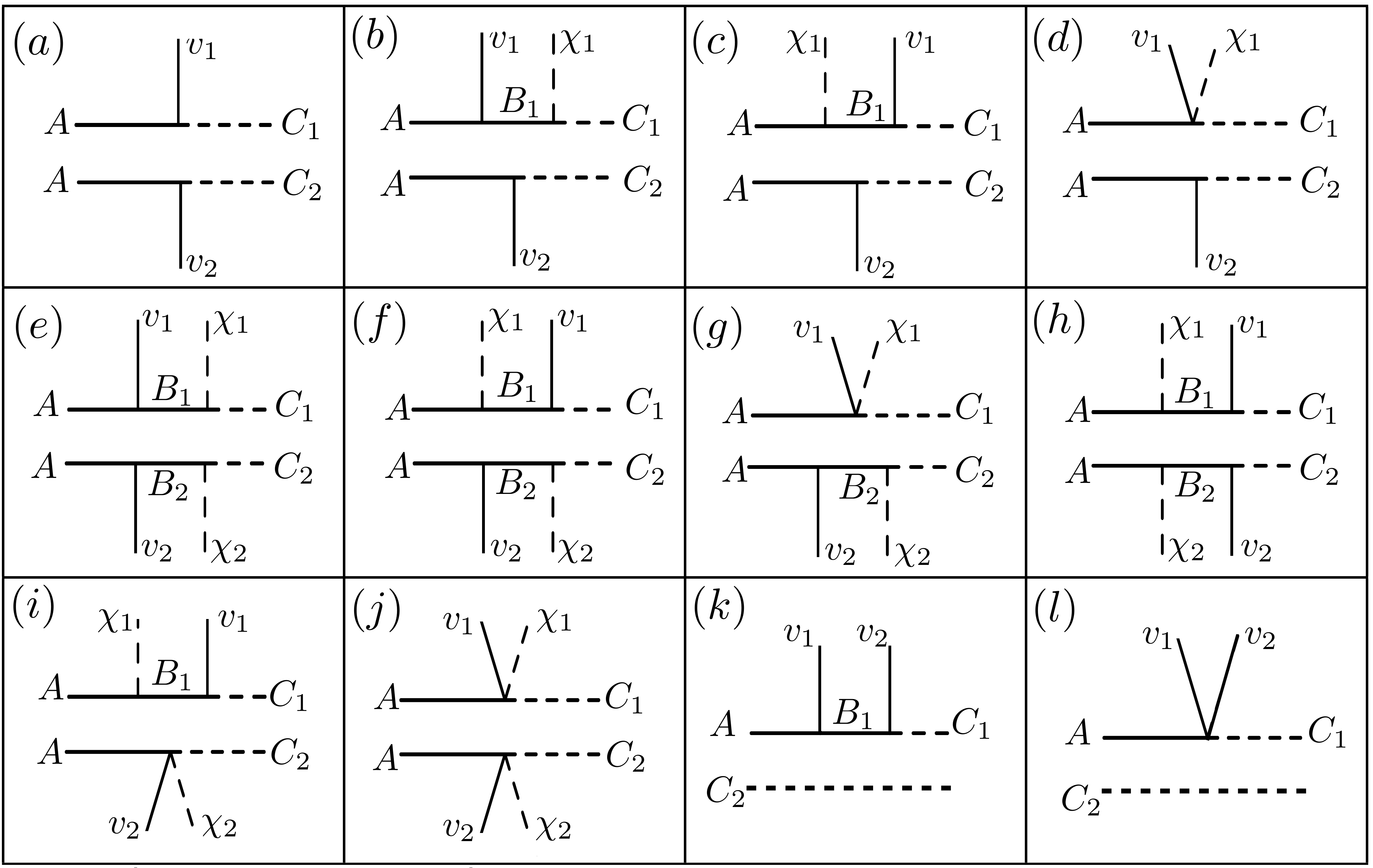}
\end{center}
\caption{Possible event topologies with two visible particles $v_1$ and $v_2$ in the final state
and up to 2 invisible particles $\left\{C_i,\chi_i\right\}$ per decay chain. $B_1$ and $B_2$ are
intermediate on-shell resonances: $M_A>M_{B_i}>M_{C_i}$.
\label{fig:diagrams} }
\end{figure}

The main goal of this paper is to illustrate the application of the Oxbridge $M_{T2}$ variable
in the more general situations described in (c-f) above. We shall be particularly concerned with
the issue (e) of multiple invisible particles and the related problem (f) of the unknown decay topology.
It is quite difficult to construct a kinematic variable (or more generally, some kind of object)
which would single out the correct event topology. An initial attempt in this direction was made in
Ref.\,\cite{Cho:2012er} which considered the decay of a {\em single} resonance $A$
(as opposed to pair production as in Fig.\,\ref{fig:example}) and studied the 
invariant mass distribution $M_V$ of the visible decay products. It was suggested that there exists
a correlation between the peak, the curvature at the peak, and the endpoint of the 
$M_V$ distribution, which in favorable scenarios can be used to infer the decay topology 
and the number of invisibles $N_{inv}$. However, this method requires $N_{vis}\ge 2$ 
and cannot be used to discriminate the topologies of Fig.\,\ref{fig:diagrams}(a-j), for
which $N_{vis}=1$ and $M_V=M_{v_i}=const$. 

It appears, therefore, that we need to study event topologies on a case by case basis,
i.e., assume in turn each one of the event topologies from Fig.\,\ref{fig:diagrams}, then test 
for consistency with the data. Note that the event topologies differ from each other in 
two aspects: 
\begin{enumerate}
\item The number of invisible particles $N_{inv}$. For example, 
the diagrams in Fig.\,\ref{fig:diagrams}(a,k,l) have $N_{inv}=N'_{inv}=1$,
the diagrams in Fig.\,\ref{fig:diagrams}(e-j) have $N_{inv}=N'_{inv}=2$,
while the diagrams in Fig.\,\ref{fig:diagrams}(b-d) have $N_{inv}=2$ and $N'_{inv}=1$.
\item Different number of intermediate on-shell resonances $B_i$.
For example, the diagrams in Fig.\,\ref{fig:diagrams}(a,d,j,l) have no intermediate resonances,
the diagrams in Fig.\,\ref{fig:diagrams}(b,c,g,i,k) have one, 
while the diagrams in Fig.\,\ref{fig:diagrams}(e,f,h) have two.
\end{enumerate}
As already discussed earlier, both of these effects jeopardize the conventional $M_{T2}$
approach, and some modifications are required. In principle, there are two possible solutions.
The first, presented in Section~\ref{sec:3plus1}, is to modify the conventional $M_{T2}$ definition
on a case by case basis --- depending on the assumed event topology. The new definitions
will use the appropriate number of invisible particles, and will also take into account the relevant 
on-shell kinematic constraints --- one constraint for each intermediate resonance $B_i$. 
Because the mass shell constraints are $3+1$-dimensional relations, this approach 
does not use the transverse $M_{T2}$ variable per se, but the alternative $3+1$ dimensional
invariant mass variable $M_2$ from \cite{Barr:2011xt}, supplemented with the appropriate 
mass-shell constraints (see also \cite{Ross:2007rm,Barr:2008hv} and the analogous 
constrained variables $M_{2C}$ and $M_{3C}$). 
The advantage of this approach is that we do not have to change the {\em interpretation} of the corresponding
kinematic endpoint $M_2^{max}$ --- it still provides a {\em saturated} bound on the parent 
mass in terms of the masses of the invisible daughter particles, similarly to eq.\,(\ref{MT2maxsaturated}).

On the other hand, from a practical standpoint, the constrained $M_2$ method has a certain disadvantage:
one has to define a new $M_2$ variable for each assumed event topology. In the majority of the cases,
those variables are {\em not} equivalent to the conventional $M_{T2}$, and therefore cannot be 
calculated with the existing $M_{T2}$ codes\,\cite{MT2library,Cheng:2008hk}. Since analytical formulas are
also unavailable, one would have to write a separate code for each variable. The mass-shell constraints 
represent an additional complication, since most likely they will have to be solved analytically 
first\,\cite{Webber:2009vm,HarlandLang:2011ih,HarlandLang:2012gn}. This is why in 
Section~\ref{sec:effective} we present an alternative approach whereby one keeps the original
definition of $M_{T2}$ (thus being able to recycle the existing numerical codes), and instead
modifies the {\em interpretation} of the kinematic endpoint $M_{T2}^{max}$ on a case by case basis.
The basic idea was raised in \cite{Polesello:2009rn}, where it was applied to the $M_{CT}$ variable\,\cite{Tovey:2008ui} in the example of chargino decay. Here we show that one can apply a similar treatment
to the $M_{T2}$ kinematic endpoints, and we also provide the recipe for a general event topology.

In Section~\ref{sec:DLSP} and Section~\ref{sec:DP} we extend our method to the case of different 
children particles ($C_1$ and $C_2$) and different parent particles ($A_1$ and $A_2$), respectively. 
Finally, in Section~\ref{sec:shapes} 
we study the impact of the number of invisible particles $N_{inv}$ on the shape of the $M_{T2_\perp}$
distribution\,\cite{Konar:2009wn}. The technique was originally proposed in \cite{Giudice:2011ib},
which focused on the shape of the $M_{T2}$ variable. Here we prefer to consider the doubly 
projected version $M_{T2_\perp}$, which is less sensitive to extraneous factors like spin correlations, 
the underlying partonic CM energy $\sqrt{s}$, etc.\,\cite{Edelhauser:2012xb}.
Thus our analysis completes the generalization of $M_{T2}$ along the lines (c-f) discussed above.
Since the directions (a) and (b) have already been discussed extensively in the literature,
for the sake of clarity and simplicity, throughout the paper we shall use the simplest assumption
$N_{vis}=N'_{vis}=1$ and we shall introduce upstream $P_T$ (i.e.,~transverse momentum for the
parent system $(A, A)$) only as necessary.

\section{Maximally constrained invariant mass variables}
\label{sec:3plus1}

In this section we revisit the event topologies from Fig.\,\ref{fig:diagrams}
and for each one define the appropriate invariant mass variable 
which provides the maximal bound on the parent mass $M_A$\,\cite{Barr:2011xt}.

\underline{The event topology of Fig.\,\ref{fig:diagrams}(a).} This is the classic $M_{T2}$ 
event topology from Fig.\,\ref{fig:example}(b), so the relevant variable is simply $M_{T2}$\,\cite{Lester:1999tx}
(or its asymmetric version if the children have different masses\,\cite{Konar:2009qr}). 
For future reference, it is convenient to introduce here the $3+1$-dimensional
version of $M_{T2}$, denoted simply as $M_2$\footnote{Not to be confused with the 
wino mass parameter in supersymmetry.}, which will make it very easy to 
incorporate additional mass-shell constraints later on. 

$M_2$ is defined on a case by case basis, following the general recipe outlined in \cite{Barr:2011xt}.
The particles in the final state are divided into two groups\footnote{The division into two groups
is motivated by the two parent hypothesis. If we do not make this assumption and instead treat the
event as a whole, we are led to the global inclusive variable $\sqrt{\hat{s}}_{min}$\,\cite{Konar:2008ei,Konar:2010ma}, which in the language of \cite{Barr:2011xt} is denoted as 
$M_1$.} (hence the subscript ``2'' on $M_2$),
one for each parent. Then, the larger of the two parent masses is minimized over the momenta 
of the invisible particles. The minimization is performed subject to all existing kinematic constraints 
and assumptions. The measurement of the missing transverse momentum $\mpt$
always provides one constraint on the net sum of the invisible transverse momenta,
and another constraint is due to the assumption that the parents have equal masses
(see discussion of item (d) in the Introduction).

In the case of the diagram in Fig.\,\ref{fig:diagrams}(a), there are no further constraints, and we get
\bea
M_{2(a)}^2(\tilde M_{C_1}; \tilde M_{C_2}) &=& \min_{P_{C_1},P_{C_2}}
\left\{ (P_{v_1}+P_{C_1})^2\right\} 
\label{M2a}\\
(P_{v_1}+P_{C_1})^2 &=& (P_{v_2}+P_{C_2})^2   \nonumber  \\
                 P_{C_1}^2 &=& \tilde M_{C_1}^2   \nonumber \\
                 P_{C_2}^2 &=& \tilde M_{C_2}^2   \nonumber \\
\vec{P}_{TC_1} + \vec{P}_{TC_2} &=& \mpt  \nonumber
\eea
Upon performing the minimization over the longitudinal momenta 
$P_{zC_1}$ and $P_{zC_2}$, one finds that eq.\,(\ref{M2a}) is equivalent\,\cite{Ross:2007rm,Barr:2011xt} 
to the usual $M_{T2}$ when the children's masses are taken to be the same
\beq
M_{2(a)}(\tilde M_{C}; \tilde M_{C}) = M_{T2}(\tilde M_{C}),
\eeq
or to the asymmetric $M_{T2}$\,\cite{Konar:2009qr}
\beq
M_{2(a)}(\tilde M_{C_1}; \tilde M_{C_2}) = M_{T2}(\tilde M_{C_1},\tilde M_{C_2}),
\label{M2aeqMT2}
\eeq
if the children's masses are kept different.

\underline{The event topology of Fig.\,\ref{fig:diagrams}(b).} 
Proceeding as before, for the diagram in Fig.\,\ref{fig:diagrams}(b) we get
\bea
M^2_{2(b)}(\tilde M_{\chi_1}, \tilde M_{C_1}; \tilde M_{C_2}) &=& \min_{P_{\chi_1},P_{C_1},P_{C_2}}
\left\{(P_{v_1}+P_{\chi_1}+P_{C_1})^2\right\} .
\label{M2b}\\
(P_{v_1}+P_{\chi_1}+P_{C_1})^2 &=& (P_{v_2}+P_{C_2})^2   \nonumber  \\
              P_{\chi_1}^2 &=& \tilde M_{\chi_1}^2   \nonumber \\
                 P_{C_1}^2 &=& \tilde M_{C_1}^2   \nonumber \\
                 P_{C_2}^2 &=& \tilde M_{C_2}^2   \nonumber \\
(P_{\chi_1}+P_{C_1})^2 &=& M_{B_1}^2 \nonumber \\
\vec{P}_{T\chi_1} + \vec{P}_{TC_1} + \vec{P}_{TC_2} &=&  \mpt   \nonumber  
\eea
With $P_{B_1}\equiv P_{\chi_1}+P_{C_1}$, this can be equivalently written as
\bea
M_{2(b)}^2(\tilde M_{\chi_1}, \tilde M_{C_1}; \tilde M_{C_2}) &=& \min_{P_{B_1},P_{C_2}}
\left\{ (P_{v_1}+P_{B_1})^2\right\} .
\label{M2bRewrite}\\
(P_{v_1}+P_{B_1})^2 &=& (P_{v_2}+P_{C_2})^2   \nonumber  \\
                 P_{B_1}^2 &=& M_{B_1}^2   \nonumber \\
                 P_{C_2}^2 &=& \tilde M_{C_2}^2   \nonumber \\
\vec{P}_{TB_1} + \vec{P}_{TC_2} &=& \mpt  \nonumber
\eea
From here, comparing to eq.\,(\ref{M2a}) and taking into account eq.\,(\ref{M2aeqMT2}), it follows that
\beq
M_{2(b)}(\tilde M_{\chi_1}, \tilde M_{C_1}; \tilde M_{C_2}) = M_{2(a)}(M_{B_1} ;\tilde M_{C_2}) 
= M_{T2}(M_{B_1},\tilde M_{C_2}).
\label{M2beqMT2}
\eeq
As expected, the kinematics of Fig.\,\ref{fig:diagrams}(b) is described by the 
asymmetric $M_{T2}$ variable\,\cite{Konar:2009qr}, with the intermediate invisible resonance $B_1$ treated 
effectively as a final state invisible particle.

\underline{The event topology of Fig.\,\ref{fig:diagrams}(c).} We get
\bea
M^2_{2(c)}(\tilde M_{\chi_1}, \tilde M_{C_1}; \tilde M_{C_2}) &=& \min_{P_{\chi_1},P_{C_1},P_{C_2}}
\left\{(P_{v_1}+P_{\chi_1}+P_{C_1})^2\right\} .
\label{M2c}\\
(P_{v_1}+P_{\chi_1}+P_{C_1})^2 &=& (P_{v_2}+P_{C_2})^2   \nonumber  \\
              P_{\chi_1}^2 &=& \tilde M_{\chi_1}^2   \nonumber \\
                 P_{C_1}^2 &=& \tilde M_{C_1}^2   \nonumber \\
                 P_{C_2}^2 &=& \tilde M_{C_2}^2   \nonumber \\
(P_{v_1}+P_{C_1})^2 &=& M_{B_1}^2 \nonumber \\
\vec{P}_{T\chi_1} + \vec{P}_{TC_1} + \vec{P}_{TC_2} &=&  \mpt   \nonumber  
\eea
This variable is new --- it cannot be related to existing versions of the $M_{T2}$ variables
as in eqs.\,(\ref{M2aeqMT2}, \ref{M2beqMT2}).

\underline{The event topology of Fig.\,\ref{fig:diagrams}(d).} Here there are no mass-shell constraints and we
have
\bea
M^2_{2(d)}(\tilde M_{\chi_1}, \tilde M_{C_1}; \tilde M_{C_2}) &=& \min_{P_{\chi_1},P_{C_1},P_{C_2}}
\left\{(P_{v_1}+P_{\chi_1}+P_{C_1})^2\right\} .
\label{M2d}\\
(P_{v_1}+P_{\chi_1}+P_{C_1})^2 &=& (P_{v_2}+P_{C_2})^2   \nonumber  \\
              P_{\chi_1}^2 &=& \tilde M_{\chi_1}^2   \nonumber \\
                 P_{C_1}^2 &=& \tilde M_{C_1}^2   \nonumber \\
                 P_{C_2}^2 &=& \tilde M_{C_2}^2   \nonumber \\
\vec{P}_{T\chi_1} + \vec{P}_{TC_1} + \vec{P}_{TC_2} &=&  \mpt   \nonumber  
\eea
It is easy to show that the minimization in eq.\,(\ref{M2d}) selects the collinear momentum configuration\,\cite{Konar:2008ei,Barr:2011xt}
\beq
\vec{P}_{\chi_1} = \frac{\tilde M_{\chi_1}}{\tilde M_{C_1}}\, \vec{P}_{C_1} 
\eeq
and eq.\,(\ref{M2d}) can be rewritten as
\bea
M^2_{2(d)}(\tilde M_{\chi_1}, \tilde M_{C_1}; \tilde M_{C_2}) &=& \min_{P_{\Psi_1},P_{C_2}}
\left\{(P_{v_1}+P_{\Psi_1})^2\right\} .
\label{M2dPsi}\\
(P_{v_1}+P_{\Psi_1})^2 &=& (P_{v_2}+P_{C_2})^2   \nonumber  \\
              P_{\Psi_1}^2 &=& (\tilde M_{\chi_1}+\tilde M_{C_1})^2   \nonumber \\
                 P_{C_2}^2 &=& \tilde M_{C_2}^2   \nonumber \\
\vec{P}_{T\Psi_1} + \vec{P}_{TC_2} &=&  \mpt   \nonumber  
\eea
in terms of an effective composite invisible particle $\Psi_1$
with mass
\beq
\tilde M_{\Psi_1} \equiv \tilde M_{\chi_1}+\tilde M_{C_1}
\label{massPsi}
\eeq
and 3-momentum 
\beq
\vec{P}_{\Psi_1}\equiv \vec{P}_{\chi_1} + \vec{P}_{C_1} 
= \frac{\tilde M_{\Psi_1}}{\tilde M_{C_1}}\, \vec{P}_{C_1} 
= \frac{\tilde M_{\Psi_1}}{\tilde M_{\chi_1}}\, \vec{P}_{\chi_1} .
\label{PPsi}
\eeq
Comparing to eq.\,(\ref{M2a}), we recognize this as the asymmetric $M_{T2}$ variable in eq.\,(\ref{M2aeqMT2})\,\cite{Konar:2009qr}
\beq
M_{2(d)}(\tilde M_{\chi_1}, \tilde M_{C_1};  \tilde M_{C_2}) 
=M_{2(a)}(\tilde M_{\chi_1}+\tilde M_{C_1}; \tilde M_{C_2}) 
=M_{T2}(\tilde M_{\chi_1}+\tilde M_{C_1},\tilde M_{C_2}). 
\eeq

\underline{The event topology of Fig.\,\ref{fig:diagrams}(e).} This is our first example with
{\em four} invisible particles:
\bea
M^2_{2(e)}(\tilde M_{\chi_1}, \tilde M_{C_1}; \tilde M_{\chi_2},\tilde M_{C_2}) &=& \min_{P_{\chi_1},P_{C_1},P_{\chi_2},P_{C_2}}
\left\{(P_{v_1}+P_{\chi_1}+P_{C_1})^2\right\} .
\label{M2e}\\
(P_{v_1}+P_{\chi_1}+P_{C_1})^2 &=& (P_{v_2}+P_{\chi_2}+P_{C_2})^2   \nonumber  \\
              P_{\chi_1}^2 &=& \tilde M_{\chi_1}^2   \nonumber \\
                 P_{C_1}^2 &=& \tilde M_{C_1}^2   \nonumber \\
              P_{\chi_2}^2 &=& \tilde M_{\chi_2}^2   \nonumber \\
                 P_{C_2}^2 &=& \tilde M_{C_2}^2   \nonumber \\
(P_{\chi_1}+P_{C_1})^2 &=& M_{B_1}^2 \nonumber \\
(P_{\chi_2}+P_{C_2})^2 &=& M_{B_2}^2 \nonumber \\
\vec{P}_{T\chi_1} + \vec{P}_{TC_1} +\vec{P}_{T\chi_2} + \vec{P}_{TC_2} &=&  \mpt   \nonumber  
\eea
By introducing $P_{B_i}\equiv P_{\chi_i}+P_{C_i}, (i=1,2)$, this can be equivalently rewritten as
\bea
M^2_{2(e)}(\tilde M_{\chi_1}, \tilde M_{C_1}; \tilde M_{\chi_2},\tilde M_{C_2}) &=& \min_{P_{B_1},P_{B_2}}
\left\{(P_{v_1}+P_{B_1})^2\right\} 
\label{M2eRewrite}\\
(P_{v_1}+P_{B_1})^2 &=& (P_{v_2}+P_{B_2})^2   \nonumber  \\
P_{B_1}^2 &=& M_{B_1}^2 \nonumber \\
P_{B_2}^2 &=& M_{B_2}^2 \nonumber \\
\vec{P}_{TB_1}+ \vec{P}_{TB_2} &=&  \mpt   \nonumber  
\eea
and therefore is again reduced to the asymmetric $M_{T2}$\,\cite{Konar:2009qr}
\beq
M_{2(e)}(\tilde M_{\chi_1}, \tilde M_{C_1}; \tilde M_{\chi_2},\tilde M_{C_2}) 
= M_{2(a)}(M_{B_1}; M_{B_2}) = M_{T2}(M_{B_1},M_{B_2}).
\eeq
Notice that $M_{2(e)}$ does {\em not} depend on the test masses 
$\tilde M_{\chi_1}$, $\tilde M_{C_1}$, $\tilde M_{\chi_1}$ and $\tilde M_{C_2}$.
In fact, the two invisible decays $B_i\to \chi_i + C_i$, $(i=1,2)$,
have no observable consequences.

\underline{The event topology of Fig.\,\ref{fig:diagrams}(f).} This is similar to the previous case,
except the order in which particles $\chi_1$ and $v_1$ appear in the decay chain is reversed
\bea
M^2_{2(f)}(\tilde M_{\chi_1}, \tilde M_{C_1}; \tilde M_{\chi_2},\tilde M_{C_2}) &=& \min_{P_{\chi_1},P_{C_1},P_{\chi_2},P_{C_2}}
\left\{(P_{v_1}+P_{\chi_1}+P_{C_1})^2\right\} .
\label{M2f}\\
(P_{v_1}+P_{\chi_1}+P_{C_1})^2 &=& (P_{v_2}+P_{\chi_2}+P_{C_2})^2   \nonumber  \\
              P_{\chi_1}^2 &=& \tilde M_{\chi_1}^2   \nonumber \\
                 P_{C_1}^2 &=& \tilde M_{C_1}^2   \nonumber \\
              P_{\chi_2}^2 &=& \tilde M_{\chi_2}^2   \nonumber \\
                 P_{C_2}^2 &=& \tilde M_{C_2}^2   \nonumber \\
(P_{v_1}+P_{C_1})^2 &=& M_{B_1}^2 \nonumber \\
(P_{\chi_2}+P_{C_2})^2 &=& M_{B_2}^2 \nonumber \\
\vec{P}_{T\chi_1} + \vec{P}_{TC_1} +\vec{P}_{T\chi_2} + \vec{P}_{TC_2} &=&  \mpt   \nonumber  
\eea
Once again, the $B_2$ mass shell constraint causes the inputs 
$\tilde M_{\chi_2}$ and $\tilde M_{C_2}$ to drop out and we get
\bea
M^2_{2(f)}(\tilde M_{\chi_1}, \tilde M_{C_1}; \tilde M_{\chi_2},\tilde M_{C_2}) &=& \min_{P_{\chi_1},P_{C_1},P_{B_2}}
\left\{(P_{v_1}+P_{\chi_1}+P_{C_1})^2\right\} .
\label{M2fRewrite}\\
(P_{v_1}+P_{\chi_1}+P_{C_1})^2 &=& (P_{v_2}+P_{B_2})^2   \nonumber  \\
              P_{\chi_1}^2 &=& \tilde M_{\chi_1}^2   \nonumber \\
                 P_{C_1}^2 &=& \tilde M_{C_1}^2   \nonumber \\
                 P_{B_2}^2 &=& M_{B_2}^2   \nonumber \\
(P_{v_1}+P_{C_1})^2 &=& M_{B_1}^2 \nonumber \\
\vec{P}_{T\chi_1} + \vec{P}_{TC_1} +\vec{P}_{TB_2} &=&  \mpt   \nonumber  
\eea
This variable is also different from $M_{T2}$, but is related to $M_{2(c)}$:
\beq
M_{2(f)}(\tilde M_{\chi_1}, \tilde M_{C_1}; \tilde M_{\chi_2},\tilde M_{C_2}) 
=M_{2(c)}(\tilde M_{\chi_1}, \tilde M_{C_1}; \tilde M_{B_2}) .
\eeq

\underline{The event topology of Fig.\,\ref{fig:diagrams}(g).} This case is similar to 
Fig.\,\ref{fig:diagrams}(e), but the $B_1$ mass shell constraint is removed:
\bea
M^2_{2(g)}(\tilde M_{\chi_1}, \tilde M_{C_1}; \tilde M_{\chi_2},\tilde M_{C_2}) &=& \min_{P_{\chi_1},P_{C_1},P_{\chi_2},P_{C_2}}
\left\{(P_{v_1}+P_{\chi_1}+P_{C_1})^2\right\} .
\label{M2g}\\
(P_{v_1}+P_{\chi_1}+P_{C_1})^2 &=& (P_{v_2}+P_{\chi_2}+P_{C_2})^2   \nonumber  \\
              P_{\chi_1}^2 &=& \tilde M_{\chi_1}^2   \nonumber \\
                 P_{C_1}^2 &=& \tilde M_{C_1}^2   \nonumber \\
              P_{\chi_2}^2 &=& \tilde M_{\chi_2}^2   \nonumber \\
                 P_{C_2}^2 &=& \tilde M_{C_2}^2   \nonumber \\
(P_{\chi_2}+P_{C_2})^2 &=& M_{B_2}^2 \nonumber \\
\vec{P}_{T\chi_1} + \vec{P}_{TC_1} +\vec{P}_{T\chi_2} + \vec{P}_{TC_2} &=&  \mpt   \nonumber  
\eea
Introducing the effective particle $\Psi$ as in eqs.\,(\ref{massPsi}, \ref{PPsi}), we get
\bea
M^2_{2(g)}(\tilde M_{\chi_1}, \tilde M_{C_1}; \tilde M_{\chi_2},\tilde M_{C_2}) &=& \min_{P_{\Psi_1},P_{B_2}}
\left\{(P_{v_1}+P_{\Psi_1})^2\right\} .
\label{M2gRewrite}\\
(P_{v_1}+P_{\Psi_1})^2 &=& (P_{v_2}+P_{B_2})^2   \nonumber  \\
                 P_{\Psi_1}^2 &=& (\tilde M_{\chi_1}+\tilde M_{C_1})^2   \nonumber \\
                 P_{B_2}^2 &=& M_{B_2}^2 \nonumber \\
\vec{P}_{T\Psi_1} +\vec{P}_{TB_2} &=&  \mpt   \nonumber  
\eea
and therefore
\beq
M_{2(g)}(\tilde M_{\chi_1}, \tilde M_{C_1}; \tilde M_{\chi_2},\tilde M_{C_2}) 
= M_{2(a)}(\tilde M_{\chi_1}+\tilde M_{C_1}; M_{B_2}) = M_{T2}(\tilde M_{\chi_1}+\tilde M_{C_1},M_{B_2}).
\eeq

\underline{The event topology of Fig.\,\ref{fig:diagrams}(h).} This is another new and non-trivial case:
\bea
M^2_{2(h)}(\tilde M_{\chi_1}, \tilde M_{C_1}; \tilde M_{\chi_2},\tilde M_{C_2}) &=& \min_{P_{\chi_1},P_{C_1},P_{\chi_2},P_{C_2}}
\left\{(P_{v_1}+P_{\chi_1}+P_{C_1})^2\right\} .
\label{M2h}\\
(P_{v_1}+P_{\chi_1}+P_{C_1})^2 &=& (P_{v_2}+P_{\chi_2}+P_{C_2})^2   \nonumber  \\
              P_{\chi_1}^2 &=& \tilde M_{\chi_1}^2   \nonumber \\
                 P_{C_1}^2 &=& \tilde M_{C_1}^2   \nonumber \\
              P_{\chi_2}^2 &=& \tilde M_{\chi_2}^2   \nonumber \\
                 P_{C_2}^2 &=& \tilde M_{C_2}^2   \nonumber \\
(P_{v_1}+P_{C_1})^2 &=& M_{B_1}^2 \nonumber \\
(P_{v_2}+P_{C_2})^2 &=& M_{B_2}^2 \nonumber \\
\vec{P}_{T\chi_1} + \vec{P}_{TC_1} +\vec{P}_{T\chi_2} + \vec{P}_{TC_2} &=&  \mpt   \nonumber  
\eea
The variable $M_{2(h)}$ cannot be reduced to one of the previous variables and would have to be evaluated
separately.

An interesting variation of the $M_{2(h)}$ variable arises in the symmetric case when the intermediate on-shell
particles $B_1$ and $B_2$ are the same, with $M_{B_1}=M_{B_2}$. Then, one can replace the two mass shell
constraints for $B_1$ and $B_2$ with the requirement that the $B_1$ and $B_2$ masses are equal, but without 
specifying the actual numerical value:
\bea
M^2_{2(h)}(\tilde M_{\chi_1}, \tilde M_{C_1}; \tilde M_{\chi_2},\tilde M_{C_2}) &=& \min_{P_{\chi_1},P_{C_1},P_{\chi_2},P_{C_2}}
\left\{(P_{v_1}+P_{\chi_1}+P_{C_1})^2\right\} .
\label{M2hnew}\\
(P_{v_1}+P_{\chi_1}+P_{C_1})^2 &=& (P_{v_2}+P_{\chi_2}+P_{C_2})^2   \nonumber  \\
              P_{\chi_1}^2 &=& \tilde M_{\chi_1}^2   \nonumber \\
                 P_{C_1}^2 &=& \tilde M_{C_1}^2   \nonumber \\
              P_{\chi_2}^2 &=& \tilde M_{\chi_2}^2   \nonumber \\
                 P_{C_2}^2 &=& \tilde M_{C_2}^2   \nonumber \\
(P_{v_1}+P_{C_1})^2 &=& (P_{v_2}+P_{C_2})^2  \nonumber \\
\vec{P}_{T\chi_1} + \vec{P}_{TC_1} +\vec{P}_{T\chi_2} + \vec{P}_{TC_2} &=&  \mpt   \nonumber  
\eea
The advantage of this approach is that one does not need to know the value of $M_{B_1}=M_{B_2}$ beforehand.

\underline{The event topology of Fig.\,\ref{fig:diagrams}(i).} This is similar to the previous case, except the
$B_2$ mass shell constraint is absent:
\bea
M^2_{2(i)}(\tilde M_{\chi_1}, \tilde M_{C_1}; \tilde M_{\chi_2},\tilde M_{C_2}) &=& \min_{P_{\chi_1},P_{C_1},P_{\chi_2},P_{C_2}}
\left\{(P_{v_1}+P_{\chi_1}+P_{C_1})^2\right\} .
\label{M2i}\\
(P_{v_1}+P_{\chi_1}+P_{C_1})^2 &=& (P_{v_2}+P_{\chi_2}+P_{C_2})^2   \nonumber  \\
              P_{\chi_1}^2 &=& \tilde M_{\chi_1}^2   \nonumber \\
                 P_{C_1}^2 &=& \tilde M_{C_1}^2   \nonumber \\
              P_{\chi_2}^2 &=& \tilde M_{\chi_2}^2   \nonumber \\
                 P_{C_2}^2 &=& \tilde M_{C_2}^2   \nonumber \\
(P_{v_1}+P_{C_1})^2 &=& M_{B_1}^2 \nonumber \\
\vec{P}_{T\chi_1} + \vec{P}_{TC_1} +\vec{P}_{T\chi_2} + \vec{P}_{TC_2} &=&  \mpt   \nonumber  
\eea
With the help of the effective invisible particle $\Psi$, this can be again simplified:
\bea
M^2_{2(i)}(\tilde M_{\chi_1}, \tilde M_{C_1}; \tilde M_{\chi_2},\tilde M_{C_2}) &=& \min_{P_{\chi_1},P_{C_1},P_{\Psi_2}}
\left\{(P_{v_1}+P_{\chi_1}+P_{C_1})^2\right\} .
\label{M2iRewrite}\\
(P_{v_1}+P_{\chi_1}+P_{C_1})^2 &=& (P_{v_2}+P_{\Psi_2})^2   \nonumber  \\
              P_{\chi_1}^2 &=& \tilde M_{\chi_1}^2   \nonumber \\
                 P_{C_1}^2 &=& \tilde M_{C_1}^2   \nonumber \\
                 P_{\Psi_2}^2 &=& (\tilde M_{\chi_2} + \tilde M_{C_2})^2   \nonumber \\
(P_{v_1}+P_{C_1})^2 &=& M_{B_1}^2 \nonumber \\
\vec{P}_{T\chi_1} + \vec{P}_{TC_1} + \vec{P}_{T\Psi_2} &=&  \mpt   \nonumber  
\eea
Comparing to eq.\,(\ref{M2c}), we see that this variable is related to $M_{2(c)}$:
\beq
M_{2(i)}(\tilde M_{\chi_1}, \tilde M_{C_1}; \tilde M_{\chi_2},\tilde M_{C_2})
=M_{2(c)}(\tilde M_{\chi_1}, \tilde M_{C_1}; \tilde M_{\chi_2}+\tilde M_{C_2}).
\eeq

\underline{The event topology of Fig.\,\ref{fig:diagrams}(j).} Here there are no mass shell constraints and we have
\bea
M^2_{2(j)}(\tilde M_{\chi_1}, \tilde M_{C_1}; \tilde M_{\chi_2},\tilde M_{C_2}) &=& \min_{P_{\chi_1},P_{C_1},P_{\chi_2},P_{C_2}}
\left\{(P_{v_1}+P_{\chi_1}+P_{C_1})^2\right\} .
\label{M2j}\\
(P_{v_1}+P_{\chi_1}+P_{C_1})^2 &=& (P_{v_2}+P_{\chi_2}+P_{C_2})^2   \nonumber  \\
              P_{\chi_1}^2 &=& \tilde M_{\chi_1}^2   \nonumber \\
                 P_{C_1}^2 &=& \tilde M_{C_1}^2   \nonumber \\
              P_{\chi_2}^2 &=& \tilde M_{\chi_2}^2   \nonumber \\
                 P_{C_2}^2 &=& \tilde M_{C_2}^2   \nonumber \\
\vec{P}_{T\chi_1} + \vec{P}_{TC_1} +\vec{P}_{T\chi_2} + \vec{P}_{TC_2} &=&  \mpt   \nonumber  
\eea
Introducing two effective invisible particles $\Psi_1$ and $\Psi_2$, this reduces to 
\bea
M^2_{2(j)}(\tilde M_{\chi_1}, \tilde M_{C_1}; \tilde M_{\chi_2},\tilde M_{C_2}) &=& \min_{P_{\Psi_1},P_{\Psi_2}}
\left\{(P_{v_1}+P_{\Psi_1})^2\right\} .
\label{M2jRewrite}\\
(P_{v_1}+P_{\Psi_1})^2 &=& (P_{v_2}+P_{\Psi_2})^2   \nonumber  \\
                 P_{\Psi_1}^2 &=& (\tilde M_{\chi_1}+ \tilde M_{C_1})^2   \nonumber \\
                 P_{\Psi_2}^2 &=& (\tilde M_{\chi_2}+ \tilde M_{C_2})^2   \nonumber \\
\vec{P}_{T\Psi_1} +\vec{P}_{T\Psi_2}  &=&  \mpt   \nonumber  
\eea
Comparing to eq.\,(\ref{M2a}), we see that
\beq
M_{2(j)}(\tilde M_{\chi_1}, \tilde M_{C_1}; \tilde M_{\chi_2},\tilde M_{C_2})
= M_{2(a)}(\tilde M_{\chi_1}+ \tilde M_{C_1}; \tilde M_{\chi_2}+\tilde M_{C_2})
= M_{T2}(\tilde M_{\chi_1}+ \tilde M_{C_1},\tilde M_{\chi_2}+\tilde M_{C_2}).
\eeq

In conclusion of this section, let us summarize its main points.
We considered the ten event topologies in Fig.\,\ref{fig:diagrams}(a-j)
and defined the corresponding maximally constrained invariant mass variables,
which fell into three categories.
\begin{itemize}
\item The variable $M_{2(a)}$ and its cousins $M_{2(b)}$, $M_{2(d)}$, $M_{2(e)}$, $M_{2(g)}$, and 
$M_{2(j)}$, all of which can be computed in terms of the asymmetric $M_{T2}$:
\bea
M_{2(b)}(\tilde M_{\chi_1}, \tilde M_{C_1}; \tilde M_{C_2}) &=& M_{2(a)}(M_{B_1}; \tilde M_{C_2}) \nonumber \\
&=& M_{T2}(M_{B_1},\tilde M_{C_2})   \nonumber \\
M_{2(d)}(\tilde M_{\chi_1}, \tilde M_{C_1}; \tilde M_{C_2}) 
&=&M_{2(a)}(\tilde M_{\chi_1}+\tilde M_{C_1}; \tilde M_{C_2}) \nonumber \\
&=&M_{T2}(\tilde M_{\chi_1}+\tilde M_{C_1},\tilde M_{C_2}) \nonumber \\
M_{2(e)}(\tilde M_{\chi_1}, \tilde M_{C_1}; \tilde M_{\chi_2},\tilde M_{C_2}) 
&=& M_{2(a)}(M_{B_1}; M_{B_2}) \nonumber \\
&=& M_{T2}(M_{B_1},M_{B_2}) \nonumber \\
M_{2(g)}(\tilde M_{\chi_1}, \tilde M_{C_1}; \tilde M_{\chi_2},\tilde M_{C_2}) 
&=& M_{2(a)}(\tilde M_{\chi_1}+\tilde M_{C_1}; M_{B_2}) \nonumber \\
&=& M_{T2}(\tilde M_{\chi_1}+\tilde M_{C_1},M_{B_2}) \nonumber\\
M_{2(j)}(\tilde M_{\chi_1}, \tilde M_{C_1}; \tilde M_{\chi_2},\tilde M_{C_2})
&=& M_{2(a)}(\tilde M_{\chi_1}+ \tilde M_{C_1}; \tilde M_{\chi_2}+\tilde M_{C_2})\nonumber \\
&=& M_{T2}(\tilde M_{\chi_1}+ \tilde M_{C_1},\tilde M_{\chi_2}+\tilde M_{C_2})\nonumber
\eea
\item The variable $M_{2(c)}$ and its friends $M_{2(f)}$ and $M_{2(i)}$:
\bea
M_{2(f)}(\tilde M_{\chi_1}, \tilde M_{C_1}; \tilde M_{\chi_2},\tilde M_{C_2}) 
&=& M_{2(c)}(\tilde M_{\chi_1}, \tilde M_{C_1}; \tilde M_{B_2})  \nonumber \\
M_{2(i)}(\tilde M_{\chi_1}, \tilde M_{C_1}; \tilde M_{\chi_2},\tilde M_{C_2})
&=&M_{2(c)}(\tilde M_{\chi_1}, \tilde M_{C_1}; \tilde M_{\chi_2}+\tilde M_{C_2}) \nonumber
\eea
\item The variable $M_{2(h)}$.
\end{itemize}
The advantage of these maximally constrained invariant mass variables
is that their endpoints provide saturated bounds on the parent mass $M_A$.
In other words, we recover the main feature in eq.\,(\ref{MT2maxsaturated}) of 
$M_{T2}$ which was lost due to the presence of the mass shell constraints.
More specifically, when we choose the test masses $\tilde M_{\chi_i}$ and $\tilde M_{C_i}$
to be the true ones (denoted without a tilde), the corresponding $M_2$ kinematic endpoint 
gives the parent mass $M_A$:
\beq
M_{2(...)}^{max} (M_{\chi_i}, M_{C_i})= M_A.
\label{M2maxsaturated}
\eeq
Here the subscript $(...)$ stands for the event topology specifier: $(a), (b), .... (j)$.
This also reveals the main problem with the maximally constrained variables $M_{2(...)}$ ---
while we recovered a saturated bound in eq.\,(\ref{M2maxsaturated}), we only managed to do so
at the cost of introducing a separate $M_2$ variable for each event topology.
Furthermore, in many cases, these variables cannot be calculated with the existing
public codes. This motivates an alternative, more practical approach, which will be the 
subject of the next section. 

\section{Effective event topology and reinterpretation of the kinematic endpoint 
of the usual $M_{T2}$ variable}
\label{sec:effective}

\subsection{General setup}

\begin{figure}[t] 
\begin{center}
\includegraphics[width=9cm]{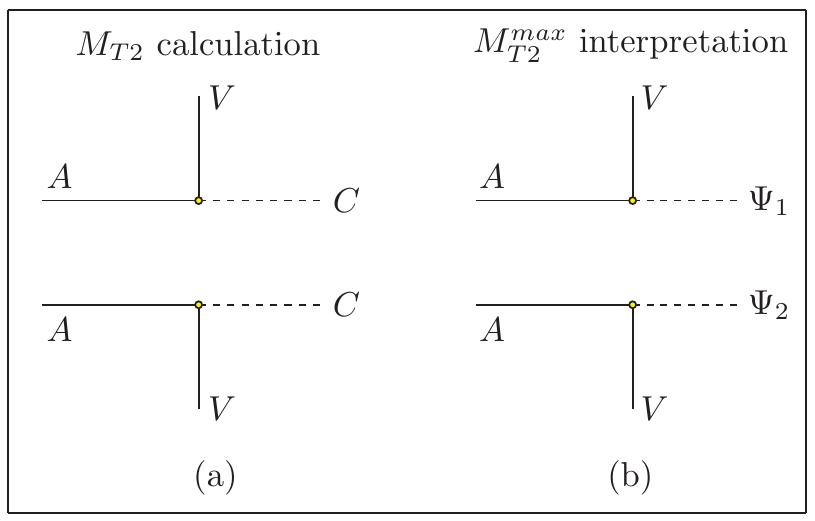}
\end{center}
\caption{The effective topology used for (a) the calculation of the $M_{T2}$ variable
and (b) the interpretation of its endpoint $M_{T2}^{max}$ in the different cases from Fig.\,\ref{fig:diagrams}.
In panel (b), $\Psi_1$ and $\Psi_2$ are effective particles whose masses are calculated in terms 
of the physical masses $M_A$, $M_B$, $M_C$ and $M_\chi$ according to the rules in Fig.\,\ref{fig:Effective}.
\label{fig:MT2max} }
\end{figure}

In this section, we revisit the 10 event topologies in Fig.\,\ref{fig:diagrams}(a-j)
and insist that we analyze all of them by means of the conventional $M_{T2}$ variable,
which is computable by the public codes. In other words, for the {\em calculation}
of the $M_{T2}$ variable, we will disregard any differences between the diagrams 
in Fig.\,\ref{fig:diagrams}(a-j)
and instead pretend that the events arise from the classic $M_{T2}$ event topology
shown in Fig.\,\ref{fig:MT2max}(a): two parents with equal masses $M_A$ are produced,
and each one subsequently undergoes a two body decay to a single visible particle 
$V$ with constant mass\footnote{For simplicity, in this paper we consider massless 
$V$ particles, i.e. $M_V=0$.} $M_V=const$ and a single invisible particle $C$ with 
mass $M_C$.  As usual, we assume that the two children particles are identical
(or at the very least, that they have a common mass $M_C$).

Following the usual procedure, we can then build the $M_{T2}(\tilde M_C)$ distribution and extract its endpoint
$M_{T2}^{max}(\tilde M_C)$. It is only at this point that we need to worry about the actual origin of the
events. In order to account for the differences between the diagrams in Fig.\,\ref{fig:diagrams}(a-j),
we propose to {\em interpret} the measured endpoint $M_{T2}^{max}(\tilde M_C)$ in terms of the
{\em effective} event topology shown in Fig.\,\ref{fig:MT2max}(b): we still produce two identical parents, 
but now they decay to {\em effective} invisible particles $\Psi_1$ and $\Psi_2$, respectively.
The interpretation of the endpoint $M_{T2}^{max}(\tilde M_C)$ is still given by the usual formula\,\cite{Cho:2007qv,Cho:2007dh}
\beq
M_{T2}^{max}(\tilde M_C) = \mu + \sqrt{\mu^2 + \tilde M_C^2},
\label{MT2maxformula}
\eeq
where
\beq
\mu\equiv\frac{M_A}{2}\left[ \left(1-\frac{M_{\Psi_1}^2}{M_A^2}\right) \left(1-\frac{M_{\Psi_2}^2}{M_A^2}\right)\right]^{1/2},
\label{mudef}
\eeq
and it is only the masses $M_{\Psi_1}$ and $M_{\Psi_2}$ of the effective particles that 
would have to be calculated on a case by case basis, depending on the assumed event 
topology from Fig.\,\ref{fig:diagrams}. The effective masses $M_{\Psi_i}, (i=1,2)$
are in general functions of the {\em true} masses of the particles appearing in the 
corresponding event topology from Fig.\,\ref{fig:diagrams}. In order to derive those
functions, we use the technique originally suggested in \cite{Polesello:2009rn}
for the case of the contransverse variable $M_{CT}$.  The main idea is to consider 
the extreme momentum configuration which gives the maximum value $M_{T2}^{max}$.
\begin{figure}[t] 
\begin{center}
\includegraphics[width=13cm]{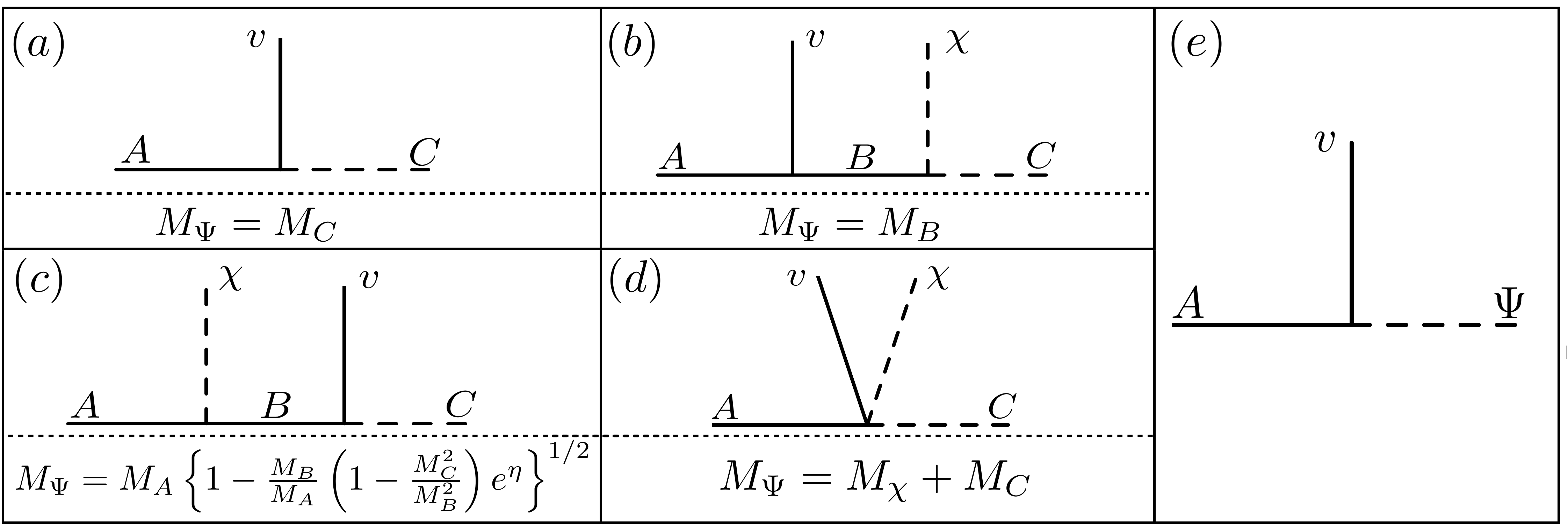}
\end{center}
\caption{Equivalence diagrams for the interpretation of $M_{T2}^{max}$ in Fig~\ref{fig:MT2max}(b).  
Each decay chain (a-d) can be replaced with the effective vertex (e). The mass $M_\Psi$
of the effective particle $\Psi$ should be suitably chosen as shown in each panel (a-d).
\label{fig:Effective} }
\end{figure}
The rules for constructing the effective mass $M_{\Psi}$ for a given cascade decay chain
are listed in Figure~\ref{fig:Effective}. Most of them should be pretty intuitive, given our discussion
in the previous section, where we already had to introduce an effective particle $\Psi$.
For example, we already encountered Fig.\,\ref{fig:Effective}(d) in eq.\,(\ref{massPsi})
and we also saw the replacement of Fig.\,\ref{fig:Effective}(b) in eq.\,(\ref{M2beqMT2}). 
Perhaps the one non-trivial example is that of Fig.\,\ref{fig:Effective}(c), where the 
effective particle mass is formed as
\beq
M_{\Psi_i}(M_A,M_{B_i},M_{C_i},M_{\chi_i})= M_A \left\{1-\frac{M_{B_i}}{M_A}\left(1-\frac{M_{C_i}^2}{M_{B_i}^2}\right) e^{\eta_i} \right\}^{1/2},
\label{eq:Meffective}
\eeq
where 
\beq
\eta_i=\cosh^{\minus 1}\left({\frac{M_A^2+M_{B_i}^2-M_{\chi_i}^2}{2 M_A M_{B_i}}}\right).
\label{eq:eta}
\eeq
When the invisible particle $\chi$ is massless ($M_\chi=0$), the effective mass 
$M_{\Psi_i}$ in eq.\,(\ref{eq:Meffective}) reduces to
\beq 
M_{\Psi_i}(M_A,M_{B_i},M_{C_i},0) =  M_{C_i}\left( \frac{M_A}{M_{B_i}}\right).   
\label{eq:cmassless}
\eeq

The advantage of the $M_{T2}$ endpoint reinterpretation approach is that one never has to stray from the
conventional procedure --- it is only at the very last stage of the analysis that the event topology issue 
comes into play. The $M_{T2}^{max}$ endpoint is always interpreted universally as in eq.\,(\ref{MT2maxformula}), 
and only the parameter $\mu$ from eq.\,(\ref{mudef}) depends on the event topology from Fig.\,\ref{fig:diagrams}:
\bea
\mu_{(a)} &=& \frac{M_A}{2}\sqrt{ \left(1-\frac{M_{C_1}^2}{M_A^2}\right) \left(1-\frac{M_{C_2}^2}{M_A^2}\right) },
\label{mua} \\
\mu_{(b)} &=& \frac{M_A}{2}\sqrt{ \left(1-\frac{M_{B_1}^2}{M_A^2}\right) \left(1-\frac{M_{C_2}^2}{M_A^2}\right) },
\label{mub} \\
\mu_{(c)} &=& e^{\eta_1/2}\,\frac{M_A}{2}\sqrt{ \frac{M_{B_1}}{M_A}\left(1-\frac{M_{C_1}^2}{M_{B_1}^2}\right) \left(1-\frac{M_{C_2}^2}{M_A^2}\right) },
\label{muc} \\
\mu_{(d)} &=& \frac{M_A}{2}\sqrt{ \left(1-\frac{(M_{C_1}+M_{\chi_1})^2}{M_A^2}\right) \left(1-\frac{M_{C_2}^2}{M_A^2}\right) },
\label{mud} \\
\mu_{(e)} &=& \frac{M_A}{2}\sqrt{ \left(1-\frac{M_{B_1}^2}{M_A^2}\right) \left(1-\frac{M_{B_2}^2}{M_A^2}\right) },
\label{mue} \\
\mu_{(f)} &=& e^{\eta_1/2}\,\frac{M_A}{2}\sqrt{ \frac{M_{B_1}}{M_A}\left(1-\frac{M_{C_1}^2}{M_{B_1}^2}\right) \left(1-\frac{M_{B_2}^2}{M_A^2}\right) },
\label{muf} \\
\mu_{(g)} &=& \frac{M_A}{2}\sqrt{ \left(1-\frac{(M_{C_1}+M_{\chi_1})^2}{M_A^2}\right) \left(1-\frac{M_{B_2}^2}{M_A^2}\right) },
\label{mug} \\
\mu_{(h)} &=& e^{(\eta_1+\eta_2)/2}\,\frac{M_A}{2}\sqrt{ \frac{M_{B_1}M_{B_2}}{M_A^2}\left(1-\frac{M_{C_1}^2}{M_{B_1}^2}\right) \left(1-\frac{M_{C_2}^2}{M_{B_2}^2}\right) },
\label{muh} \\
\mu_{(i)} &=& e^{\eta_1/2}\,\frac{M_A}{2}\sqrt{ \frac{M_{B_1}}{M_A}\left(1-\frac{M_{C_1}^2}{M_{B_1}^2}\right) \left(1-\frac{(M_{C_2}+M_{\chi_2})^2}{M_A^2}\right) },
\label{mui} \\
\mu_{(j)} &=& \frac{M_A}{2}\sqrt{ \left(1-\frac{(M_{C_1}+M_{\chi_1})^2}{M_A^2}\right) \left(1-\frac{(M_{C_2}+M_{\chi_2})^2}{M_A^2}\right) },
\label{muj}
\eea
with $\eta_i$ still given by eq.\,(\ref{eq:eta}).

\subsection{Application to chargino decays}

A well-motivated and relevant physics case illustrating the decay topologies in Fig.\,\ref{fig:Effective}
is provided by the chargino decays in supersymmetry. If we identify the parent particle
$A$ with the chargino $\tilde \chi^+$ and the visible SM particle with a charged lepton $\ell^+$, 
the invisible daughter particles could be: a neutralino $\tilde\chi^0$, a sneutrino $\tilde\nu$,
or a SM neutrino $\nu$. Then each of the decay chains in Fig.\,\ref{fig:Effective}(a-d) can be
interpreted as follows:
\begin{itemize}
\item The topology of Fig.\,\ref{fig:Effective}(a): $A\to v+C$.
This corresponds to a two-body decay of the chargino to a lepton and sneutrino:
$\tilde\chi^+\to\ell^++\tilde\nu$, with the sneutrino being the lightest supersymmetric particle (LSP). 
\item The topology of Fig.\,\ref{fig:Effective}(b): $A\to v+B$, followed by $B\to \chi+C$.
This is similar to the previous case, where we identify $B$ with the sneutrino:
$\tilde\chi^+\to\ell^++\tilde\nu$, only this time the sneutrino decays further invisibly: 
$\tilde\nu     \to\nu  +\tilde\chi^0$.
\item The topology of Fig.\,\ref{fig:Effective}(c): $A\to \chi+B$, followed by $B\to v+C$.
Here there are two possible examples. The first one is analogous to the sneutrino
decay considered above --- only this time the intermediate particle is a charged slepton 
$\tilde\ell^+$ and we get $\tilde\chi^+\to\nu+\tilde\ell^+$, followed by
$\tilde\ell^+\to\ell^++\tilde\chi^0$. The invisible particle masses are identified as
$M_C=M_{\tilde\chi ^0}$ and $M_\chi=M_{\nu}=0$. Another possibility is to have the chargino decay 
to an on-shell $W$ boson: $\tilde \chi^+\to \tilde\chi^0 + W^+$, followed by
$W^+\to \ell^+ + \nu$, in which case the invisible masses are 
$M_C=M_{\nu}=0$ and $M_\chi=M_{\tilde\chi ^0}$.
\item The topology of Fig.\,\ref{fig:Effective}(d): $A\to v+ \chi+C$.
This is realized if the chargino two-body decays are closed and 
it decays via a virtual slepton or $W$ boson: $\tilde \chi^+\to \ell^++\nu +\tilde\chi^0$.
\end{itemize}
By pairing up these 4 cases, we can obtain all 10 event topologies from Fig.\,\ref{fig:diagrams}(a-j).

As a specific scenario realizing these patterns, let us consider the 
{\tt Tchislepslep} simplified model\,\cite{SMS}, where 
$M_{\tilde\chi^+}>M_{\tilde\nu_L}\simeq M_{\tilde \ell^+_L}>M_{\tilde\chi^0}$.
The masses of the sneutrino and the charged slepton are taken to be equal because they 
belong to the same $SU(2)_W$ doublet (they are both left-handed).
The two possible chargino decays are given by Fig.\,\ref{fig:Effective}(b) and
Fig.\,\ref{fig:Effective}(c) and have equal branching fractions.
The three possible ways of pairing up Fig.\,\ref{fig:Effective}(b) and Fig.\,\ref{fig:Effective}(c)
lead to the event topologies of Fig.\,\ref{fig:diagrams}(e), Fig.\,\ref{fig:diagrams}(f)
and Fig.\,\ref{fig:diagrams}(h):
\begin{itemize}
\item {\em Sneutrino-sneutrino.} When both charginos decay through a sneutrino
as in Fig.\,\ref{fig:Effective}(b), we obtain the event topology of Fig.\,\ref{fig:diagrams}(e)
with the identifications $M_A=M_{\tilde \chi^+}$, $M_{B_1}=M_{B_2}=M_{\tilde\nu}$, 
$M_{C_1}=M_{C_2}=M_{\tilde\chi^0}$, and $M_{\chi_1}=M_{\chi_2}=M_{\nu}=0$.
Therefore the $M_{T2}$ endpoint in eq.\,(\ref{MT2maxformula}) should be interpreted with the 
$\mu_{(e)}$ parameter taken from eq.\,(\ref{mue}):
\beq
\mu_{(e)}\equiv \mu_{\tilde\nu\tilde\nu} = \frac{M_{\chi^+}}{2}\left(1-\frac{M_{\tilde\nu}^2}{M_{\tilde\chi^+}^2}\right) .
\label{musnusnu}
\eeq
\item {\em Slepton-sneutrino.} The hybrid pairing of Fig.\,
\ref{fig:Effective}(b) and
Fig.\,\ref{fig:Effective}(c) leads to the event topology of Fig.\,\ref{fig:diagrams}(f)
with the identifications $M_A=M_{\tilde \chi^+}$, $M_{B_1}=M_{\tilde\ell^+}$, $M_{B_2}=M_{\tilde\nu}$, 
$M_{C_1}=M_{C_2}=M_{\tilde\chi^0}$, and $M_{\chi_1}=M_{\chi_2}=M_{\nu}=0$.
The $M_{T2}$ endpoint for such events is interpreted with $\mu_{(f)}$ from eq.\,(\ref{muf}):
\beq
\mu_{(f)}\equiv \mu_{\tilde\ell\tilde\nu} = \frac{M_{\chi^+}}{2}
\sqrt{\left(1-\frac{M_{\tilde \chi^0}^2}{M_{\tilde \ell}^2}\right)
\left(1-\frac{M_{\tilde \nu}^2}{M_{\tilde \chi^+}^2}\right) }.
\label{muslepsnu}
\eeq
\item {\em Slepton-slepton.} When both charginos decay through a slepton
as in Fig.\,\ref{fig:Effective}(c), we obtain the event topology of Fig.\,\ref{fig:diagrams}(h)
with the identifications $M_A=M_{\tilde \chi^+}$, $M_{B_1}=M_{B_2}=M_{\tilde\ell^+}$, 
$M_{C_1}=M_{C_2}=M_{\tilde\chi^0}$, and $M_{\chi_1}=M_{\chi_2}=M_{\nu}=0$.
The $M_{T2}$ endpoint for such events is interpreted with $\mu_{(h)}$ from eq.\,(\ref{muh}):
\beq
\mu_{(h)}\equiv \mu_{\tilde\ell\tilde\ell} = \frac{M_{\chi^+}}{2}
\left(1-\frac{M_{\tilde \chi^0}^2}{M_{\tilde \ell}^2}\right).
\label{muslepslep}
\eeq
\end{itemize}

\begin{figure}[t] 
\begin{center}
\includegraphics[width=9cm]{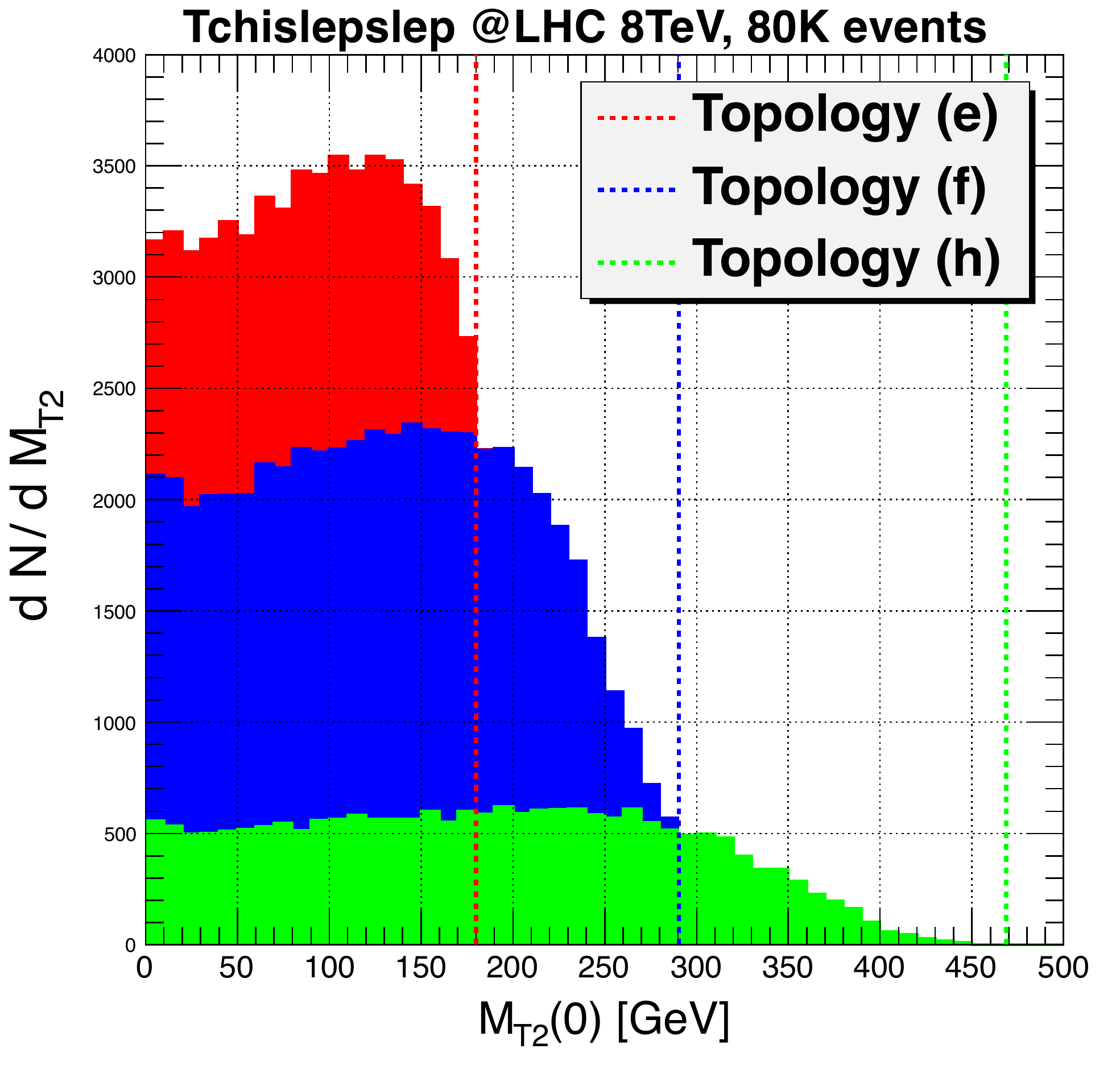} 
\end{center}
\caption{Distributions of the $M_{T2}$ variable (for zero test mass $\tilde M_C=0$)
for the {\tt Tchislepslep} simplified model, with $M_{\tilde\chi^+}=500$ GeV,
$M_{\tilde\nu_L}\simeq M_{\tilde \ell^+_L}=400$ GeV and $M_{\tilde\chi^0}=100$ GeV.
The chargino branching fractions are fixed as 
$B(\tilde\chi^+\to\ell^++\tilde\nu)= B(\tilde\chi^+\to\nu+\tilde\ell^+) =50\%$. 
Results are shown for direct chargino pair production at the LHC with 8 TeV CM energy.
The vertical dashed lines mark the expected endpoints for each event topology:
sneutrino-sneutrino (red), sneutrino-slepton (blue) and slepton-slepton (green).
\label{fig:LM7} }
\end{figure}

The {\tt Tchislepslep} scenario is illustrated in Figure~\ref{fig:LM7}, where we show 
$M_{T2}$ distributions from Monte Carlo 
simulations for direct chargino pair production, at the LHC with 8 TeV CM energy.
The electroweak mass spectrum is chosen as $M_{\tilde\chi^+}=500$ GeV,
$M_{\tilde\nu_L}\simeq M_{\tilde \ell^+_L}=400$ GeV and $M_{\tilde\chi^0}=100$ GeV,  
while the chargino branching fractions are fixed as 
$B(\tilde\chi^+\to\ell^++\tilde\nu)= B(\tilde\chi^+\to\nu+\tilde\ell^+) =50\%$. 
For simplicity, we calculate $M_{T2}$ with zero test mass ($\tilde M_C=0$), 
so that the endpoint formula in eq.\,(\ref{MT2maxformula}) simplifies to
\beq
M_{T2}^{max}(\tilde M_C=0) = 2\mu 
= \left\{ 
\begin{array}{ll}
180\ {\rm GeV} & {\rm for\ sneutrino-sneutrino\ events,}\\
290\ {\rm GeV} & {\rm for\ slepton-sneutrino\ events,}\\
469\ {\rm GeV} & {\rm for\ slepton-slepton\ events.}\\
\end{array}   \right.
\label{MT2maxformula0}
\eeq
where the $\mu$ parameter was calculated from eqs.\,(\ref{musnusnu}-\ref{muslepslep}), depending on the 
event type. In Figure~\ref{fig:LM7}, sneutrino-sneutrino events are plotted in red, 
slepton-sneutrino events are plotted in blue and slepton-slepton events are plotted in green.
The color-coded vertical dashed lines mark the corresponding $M_{T2}$ endpoints 
expected from eqs.\,(\ref{musnusnu}-\ref{MT2maxformula0}).

We conclude this section with a discussion of the {\tt TChiww} SMS model, in which the sleptons and sneutrinos
are heavy, and both charginos decay through an on-shell $W$ boson instead
(provided, of course, that $M_{\tilde \chi^+}> M_W +M_{\tilde \chi^0}$).
The corresponding topology is the one from Fig.\,\ref{fig:diagrams}(h), 
where $C_i= \nu$ and $\chi_i=\tilde\chi^0$  is the neutralino LSP.
The effective masses from eq.\,(\ref{eq:Meffective}) are
\beq
M_{\Psi_1} =M_{\Psi_2} =M_{\tilde \chi^+}\, \sqrt{1-\frac{M_W}{M_{\tilde \chi^+}} e^{\eta}},
\eeq
where
\beq
\eta=\cosh^{\minus 1}\left(\frac{M_{\tilde \chi^+}^2+M_W^2-M_{\tilde \chi^0}^2}{2 M_{\tilde \chi^+} M_{W}}\right).
\eeq
Then the $M_{T2}$ endpoint (with zero test mass $\tilde M_{\tilde\chi^0}$) is given by
\beq
M_{T2}^{max}\left(0\right)   =M_W\,  e^{\eta}.
\eeq

Finally, if the mass splitting between the chargino and the neutralino is small 
($M_{\tilde \chi^+}-M_{\tilde \chi^0} < M_W$), the chargino decays as in
Fig.\,\ref{fig:Effective}(d). The event topology is given in Fig.\,\ref{fig:diagrams}(j)
and the $M_{T2}$ endpoint (again with zero test mass $\tilde M_{\tilde\chi^0}$) is 
\beq
M_{T2}^{max}\left(0\right)   = M_{\tilde\chi^+} \left(1-\frac{M_{\tilde \chi^0}^2}{M_{\tilde \chi^+}^2}\right).
 \eeq

\section{Reinterpretation of the kinematic endpoint of the asymmetric $M_{T2}$ variable}
\label{sec:DLSP}

In the effective topology of Fig.\,\ref{fig:MT2max}(a), the test masses for the two invisible children 
were chosen to be the same: $\tilde M_C$. However, as demonstrated in \cite{Konar:2009qr},
it is straightforward to generalize the standard $M_{T2}$ calculation in Fig.\,\ref{fig:MT2max}(a)
to allow for {\em different} test masses for the two children particles. In other words, we could use 
the effective diagram in Fig.\,\ref{fig:MT2max}(b) not only for the {\em interpretation} of the
kinematic endpoint, but also for the actual {\em calculation} of the (asymmetric) $M_{T2}$  
variable itself, in terms of two {\em different} test masses\footnote{Recall our notation that 
test masses always carry a tilde.}, $\tilde M_{\Psi_1}$ 
and $\tilde M_{\Psi_2}$\,\cite{Konar:2009qr}:
\bea
M_{T2D}{\scriptstyle(\tilde{M}_{\Psi_1}, \tilde{M}_{\Psi_2})} &\equiv&  
\min_{\scriptstyle \vec{P}_{T\Psi_1}, \vec{P}_{T\Psi_2}}
\left\{\max\left\{ M_{TA_1}{\scriptstyle(\vec{P}_{T\Psi_1},\tilde M_{\Psi_1})},\, 
                         M_{TA_2}{\scriptstyle(\vec{P}_{T\Psi_2},\tilde M_{\Psi_2})} \right\} \right\}, 
\label{eq:DLSPmt2} 
\\
{\scriptstyle\vec{P}_{T\Psi_1} + \vec{P}_{T\Psi_2}} &=& {\scriptstyle\mpt}  \nonumber
\eea
where $M_{TA_i}$ is the transverse mass of the parent particle $A_i$
\beq
M_{TA_i}{\scriptstyle(\vec{P}_{T\Psi_i},\tilde M_{\Psi_i})}
\equiv
\left[M_{V_i}^2+\tilde M_{\Psi_i}^2+ 2\left( \sqrt{M_{V_i}^2+\vec P_{TV_i}^2}
\sqrt{\tilde M_{\Psi_i}^2+\vec P_{T\Psi_i}^2}
-\vec P_{TV_i} \cdot \vec P_{T\Psi_i}\right) \right]^{1/2}.
\label{MTAi}
\eeq
and the subscript ``D'' in eq.\,(\ref{eq:DLSPmt2}) is used to remind the reader that 
the asymmetric $M_{T2}$ variable uses two different test mass inputs.
In eq.\,(\ref{MTAi}), $M_{V_i}$ and $\vec{P}_{TV_i}$ are correspondingly 
the invariant mass and transverse momentum 
of the effective visible particle $V_i$ resulting from the decay of $A_i$. Similarly, 
$\tilde M_{\Psi_i}$ and $\vec P_{T\Psi_i}$ are the test mass and test transverse momentum of the 
effective invisible particle $\Psi_i$ (see Fig.\,\ref{fig:MT2max}(b)).
As usual, $\vec P_{T\Psi_1}$ and $\vec P_{T\Psi_2}$ are subject to the missing transverse 
momentum constraint, and then varied to find the minimum of the function in eq.\,(\ref{eq:DLSPmt2}).

When the visible particles are massless ($M_{V_i}=0$), as we are considering here,
the kinematic endpoint of $M_{T2D}$ is given by \cite{Konar:2009qr}
\beq
M_{T2D}^{max}{\scriptstyle\left(\tilde{M}_{\Psi_1}, \tilde{M}_{\Psi_2}\right)}=
\left[\left( \mu+ 
\sqrt{ \mu^2 +\tilde M_{+}^2+\frac{\tilde M_{-}^4}{4 \mu^2}}\right)^2-\frac{\tilde M_{-}^4}{4 \mu^2}\right]^{1/2} ,
 \label{eq:dlsp}
 \eeq
 where the parameter $\mu$ encoding the dependence on the physical masses is still given by eq.\,(\ref{mudef}), while
 the two test masses enter through the combinations
\bea
\tilde M_{+}^2 &=& \frac{1}{2}\left(\tilde M_{\Psi_1}^2+\tilde M_{\Psi_2}^2\right), \\
\tilde M_{-}^2  &=& \frac{1}{2}\left|\tilde M_{\Psi_1}^2-\tilde M_{\Psi_2}^2\right|. 
\eea

Obviously, one can always go from the asymmetric $M_{T2D}$ variable in eq.\,(\ref{eq:DLSPmt2}) 
back to the original $M_{T2}$ variable\,\cite{Lester:1999tx}, simply by setting the two test masses to be equal: 
\beq
M_{T2D}{\scriptstyle\left(\tilde{M}_{\Psi}, \tilde{M}_{\Psi}\right)}  = M_{T2}{\scriptstyle\left(\tilde{M}_{\Psi}\right)} .
\eeq
Furthermore, the $M_{T2D}$ endpoint in eq.\,(\ref{eq:dlsp}) only allows us to measure one parameter: $\mu$, and 
the same can be said about the usual $M_{T2}$ endpoint in eq.\,(\ref{MT2maxformula}) as well.
A natural question then is whether there is any benefit at all from introducing the more complicated
variable $M_{T2D}$. We see two motivations for considering $M_{T2D}$:
\begin{itemize}
\item In the context of manifestly asymmetric event topologies like the ones shown in 
Fig.\,\ref{fig:diagrams}(b), Fig.\,\ref{fig:diagrams}(c), Fig.\,\ref{fig:diagrams}(d), Fig.\,\ref{fig:diagrams}(f),
Fig.\,\ref{fig:diagrams}(g) and Fig.\,\ref{fig:diagrams}(i), the language of $M_{T2D}$ is more appropriate
because the masses of the {\em effective} invisible particles $\Psi_1$ and $\Psi_2$ are different, even if the 
masses of the particles $C_1$ and $C_2$ at the end of the decay chains are the same.
\item More importantly, by considering the asymmetric $M_{T2D}$ variable, one could 
in principle find not just one constraint among the three unknown parameters 
$M_A$, $M_{\Psi_1}$ and $M_{\Psi_2}$, but the actual values of all three parameters themselves\,\cite{Barr:2009jv,Konar:2009qr}. To this end, one needs to consider events in which the 
parent $AA$ system recoils against upstream objects (from initial state radiation or prior decays)
with a net upstream transverse momentum $U_T$. One then measures the endpoint 
in eq.\,(\ref{eq:dlsp}) and compares the results in different $U_T$ bins. In general, the 
results for the $M_{T2D}$ endpoint will depend on the value of $U_T$. 
However, when the test masses $\tilde M_{\Psi_i}$ are equal to the corresponding
{\em true values} $M_{\Psi_i}$, the $U_T$ dependence disappears
and $M_{T2D}^{max}$ becomes independent of the upstream momentum\,\cite{Konar:2009qr}:
\beq
\left. \frac{\partial M_{T2D}^{max}}{\partial U_T}\right|_{\tilde M_{\Psi_i}=M_{\Psi_i}}=0.
\label{eq:noUTdependence}
\eeq  
The true location of the invisible masses $M_{\Psi_i}$ is often also revealed as the 
crossing point of several creases in the two-dimensional hyper-surface defined by the 
function $M_{T2D}^{max}{\scriptstyle\left(\tilde{M}_{\Psi_1}, \tilde{M}_{\Psi_2}\right)}$\,\cite{Barr:2009jv}.
\end{itemize}

\section{Interpretation in the case of different parent particles} 
\label{sec:DP}

\begin{figure}[t] 
\begin{center}
\includegraphics[width=9cm]{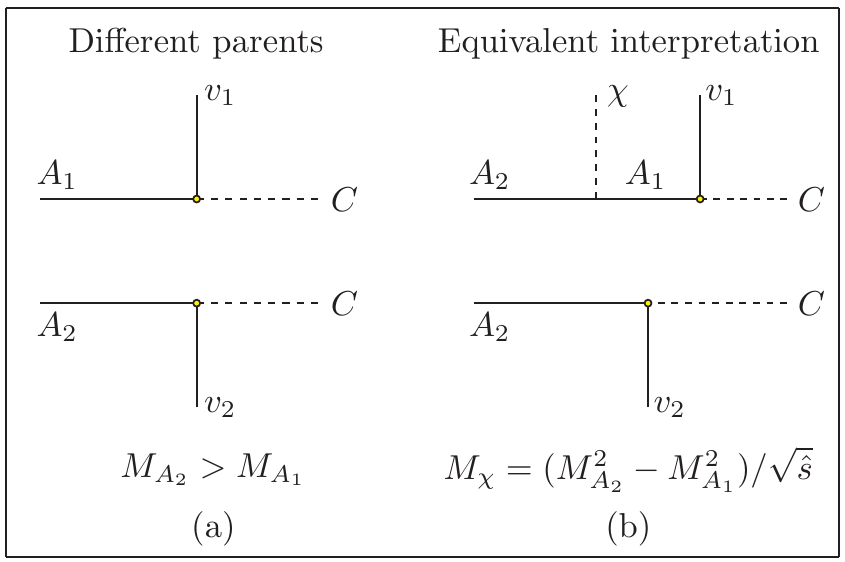}
\end{center}
\caption{(a) A generic event topology with different parent particles and (b) the effective event topology 
for its interpretation as discussed in the text.
\label{fig:DP} }
\end{figure}

As another application of the effective topology method, in this section we revisit the case 
of {\em different} parent particles $\left(M_{A_1} \neq M_{A_2}\right)$ as illustrated in Fig.\,\ref{fig:DP}(a). 
A well-motivated example of such a ``coproduction" channel is provided by associated
gluino-squark production which was studied in \cite{Nojiri:2008hy,Nojiri:2008vq}, finding a correlation 
between the endpoint $M_{T2}^{max}$ of the conventional $M_{T2}$ variable and 
the larger of the two parent masses:
\beq
M_{T2}^{(max)}(\tilde M_C=M_C) = \max\left(M_{A_1}, M_{A_2}\right).
\label{MT2maxDP}
\eeq  
This result, also suggested in \cite{Barr:2009wu}, is the analogue of eq.\,(\ref{MT2maxsaturated})
for the case of different parents.
Here we point out that the relation in eq.\,(\ref{MT2maxDP}) does not hold in general. 
The main result in this section will be the proper interpretation of the $M_{T2}$ kinematic endpoint
in the case of different parents (Fig.\,\ref{fig:DP}(a)). The discussion will be organized as follows:
in Section~\ref{sec:UTzero} we first treat the case with no upstream momentum ($U_T=0$)
and then in Section~\ref{sec:UT} we discuss the more general case with $U_T\ne 0$.

\subsection{Events with no upstream momentum ($U_T=0$)}
\label{sec:UTzero}

We start from the known fact that the $M_{T2}$ endpoint is invariant under ``back-to-back" boosts
in the transverse plane\,\cite{Cho:2007dh}. In the conventional case of Fig.\,\ref{fig:example}(b) with
identical parents, this boost brings both parent particles to rest. However, when the parents have 
different masses as in Fig.\,\ref{fig:DP}(a), no back-to-back boost can bring the parents to rest 
simultaneously, and the ``back-to-back" boost invariance of $M_{T2}^{max}$ will be lost.
More specifically, if the parent system $(A_1, A_2)$ is produced with some total CM energy $\sqrt{\hat s}$,
the parent particles $A_i$ are boosted by the respective factors
\bea
\textstyle
\eta_1{\scriptstyle \left(\sqrt{\hat s}\right)} &=&\cosh^{\minus 1}\left({\frac{\hat s+M_{A_1}^2-M_{A_2}^2}{2\hat s\, M_{A_1}}}\right),   \label{eta1}\\
\eta_2{\scriptstyle \left(\sqrt{\hat s}\right)} &=&\cosh^{\minus 1}\left({\frac{\hat s+M_{A_2}^2-M_{A_1}^2}{2\hat s\, M_{A_2}}}\right).   \label{eta2}
\eea
For definiteness and without loss of generality, from now on we shall assume that $M_{A_2}>M_{A_1}$, 
so that  $\max\left(M_{A_1}, M_{A_2}\right)=M_{A_2}$. Then from eqs.\,(\ref{eta1}, \ref{eta2}) it also follows that
$\eta_2 < \eta_1$. Thus if we perform a ``back-to-back'' boost on the particles $A_1$ and $A_2$ with the 
boost factor $\eta_2$ corresponding to the heavier particle, the result will be that particle $A_2$ will be 
brought to rest, while the lighter parent particle $A_1$ will have some residual boost 
\beq
{\delta \eta
{\scriptstyle \left(\sqrt{\hat s}\right)}} 
\equiv \eta_1{\scriptstyle \left(\sqrt{\hat s}\right)}-\eta_2{\scriptstyle \left(\sqrt{\hat s}\right)}
= \cosh^{\minus 1} \left[ \frac{M_{A_2}^2+M_{A_1}^2- \left(\frac{M_{A_2}^2-M_{A_1}^2}{\sqrt{\hat s}}\right)^2}{2 M_{A_2} M_{A_1} }\right]\, .
\label{eq:gapeta}
\eeq
Comparing this to eq.\,(\ref{eq:eta}), we see that eq.\,(\ref{eq:eta}) and eq.\,(\ref{eq:gapeta}) become identical if we 
identify
\beq
M_{\chi}{\scriptstyle \left(\sqrt{\hat s}\right)} \equiv
\frac{M_{A_2}^2-M_{A_1}^2}{\sqrt{\hat s}}  .
\label{eq:MchiInt}
\eeq
The physical meaning of eq.\,(\ref{eq:MchiInt}) is the following --- the events in Fig.\,\ref{fig:DP}(a), which 
represent pair production of {\em different} parent particles, are instead reinterpreted as in Fig.\,\ref{fig:DP}(b),
which shows the pair production of {\em identical} parents $A_2$, one of which decays into $A_1$ plus a hypothetical
invisible particle $\chi$ whose mass is given by eq.\,(\ref{eq:MchiInt}). At this point it is important to note that
the mass $M_{\chi}$ as defined in eq.\,(\ref{eq:MchiInt}) is not constant, but carries dependence on the CM 
energy $\sqrt{\hat{s}}$, which is inherited from eqs.\,(\ref{eta1}, \ref{eta2}). At hadron colliders, events 
are produced with varying $\sqrt{\hat{s}}$, thus in general $M_{\chi}$ takes values in the interval
\beq
0 \le M_{\chi}{\scriptstyle \left(\sqrt{\hat s}\right)} \leq M_{A_2}-M_{A_1} \label{eq:chiRange}\, ,
\eeq
where the lower bound corresponds to $\sqrt{\hat s} \rightarrow \infty$ 
and the upper bound is obtained at threshold $\sqrt{\hat s} = M_{A_1}+M_{A_2}$. 

Once we have an event topology with equal parents as in Fig.\,\ref{fig:DP}(b), we already know
how to interpret the $M_{T2}$ endpoint --- we just follow the prescription from Section~\ref{sec:effective}.
The topology of Fig.\,\ref{fig:DP}(b) is of the type shown in Fig.\,\ref{fig:diagrams}(c), so for
the upper decay chain we need to introduce an effective invisible particle $\Psi_1$
as prescribed in Fig.\,\ref{fig:Effective}(c) and eqs.\,(\ref{eq:Meffective}, \ref{eq:eta}):
 \bea
M_{\Psi_1}&=& M_{A_2} \left\{1-\frac{M_{A_1}}{M_{A_2}} \left(1-\frac{M_C^2}{M_{A_1}^2}\right) e^{\eta{\scriptstyle \left(\sqrt{\hat s}\right)}}\right\} \, , \\
\eta{\scriptstyle \left(\sqrt{\hat s}\right)}&=&\cosh^{\minus 1} \left\{ \frac{M_{A_2}^2+M_{A_1}^2-M_\chi^2{\scriptstyle \left(\sqrt{\hat s}\right)}}{2 M_{A_2} M_{A_1}}\right\}  = \delta \eta {\scriptstyle \left(\sqrt{\hat s}\right)}.
\eea
Then, the kinematic endpoint of the usual $M_{T2}$ variable is found from eq.\,(\ref{MT2maxformula}):
\beq
M_{T2}^{max}({\scriptstyle \sqrt{\hat s}}, {\scriptstyle\tilde M_C}) = 
\mu{\scriptstyle \left(\sqrt{\hat s}\right)}+\sqrt{\mu{\scriptstyle \left(\sqrt{\hat s}\right)}^2+\tilde M^2_C}\, ,
\label{eq:MT2maxDP}
\eeq
with $ \mu{\scriptstyle \left(\sqrt{\hat s}\right)}$ given by eq.\,(\ref{muc})
\beq
\mu {\scriptstyle \left(\sqrt{\hat s}\right)}  = e^{{\eta {\scriptstyle \left(\sqrt{\hat s}\right)}}/2} \cdot \frac{M_{A_2}}{2}
\sqrt{\frac{M_{A_1}}{M_{A_2}}\left(1-\frac{M_C^2}{M_{A_1}^2}\right)\left(1-\frac{M_C^2}{M_{A_2}^2}\right)}.
\label{eq:mueta}
\eeq

Notice that the kinematic endpoint $M_{T2}^{max}$ in eq.\,(\ref{eq:MT2maxDP}) this time
depends not only on the test child mass $\tilde M_C$, but also on the partonic CM energy $\sqrt{\hat s}$.
At the LHC, $\sqrt{\hat s}$ is not constant, but varies from one event to another in 
accordance with the parton distribution functions (PDFs). Therefore, the kinematic endpoint in eq.\,(\ref{eq:MT2maxDP}) will in general\footnote{Modulo the special case where $A_1$ and $A_2$ 
are produced in the decay of some heavy narrow resonance $X$ as $X\to A_1 A_2$\,\cite{Han:2009ss,Edelhauser:2012xb}.} get smeared. 
As we can see from eqs.\,(\ref{eq:gapeta}, \ref{eq:mueta}), $\mu {\scriptstyle \left(\sqrt{\hat s}\right)}$ 
is an increasing function of $\sqrt{\hat s}$. Since $\sqrt{\hat s}$ itself varies from its threshold value
$\sqrt{\hat s}=M_{A_1}+M_{A_2}$ to $\sqrt{\hat s}\to \infty$, the function $\mu {\scriptstyle \left(\sqrt{\hat s}\right)}$
takes values in
\beq
\mu_{min} \le  \mu {\scriptstyle \left(\sqrt{\hat s}\right)} \le \mu_{max},
\eeq
where
\bea
\mu_{min} &=& \lim_{\scriptstyle \sqrt{\hat s}\to M_{A_1}+M_{A_2}} \mu {\scriptstyle \left(\sqrt{\hat s}\right)} =
\frac{M_{A_2}}{2}\sqrt{\frac{M_{A_1}}{M_{A_2}}\left(1-\frac{M_C^2}{M_{A_1}^2}\right)\left(1-\frac{M_C^2}{M_{A_2}^2}\right)},
\label{eq:mumin}
\\
\mu_{max} &=& \lim_{\scriptstyle \sqrt{\hat s}\to \infty} \mu {\scriptstyle \left(\sqrt{\hat s}\right)} =
\frac{M_{A_2}}{2}\sqrt{\left(1-\frac{M_C^2}{M_{A_1}^2}\right)\left(1-\frac{M_C^2}{M_{A_2}^2}\right)} 
= \mu_{min}\, \sqrt{\frac{M_{A_2}}{M_{A_1}}}.
\label{eq:mumax}
\eea
Now it follows from eq.\,(\ref{eq:MT2maxDP}) that even if we choose the test mass $\tilde M_C$ to be
the true mass $M_C$, the corresponding $M_{T2}$ kinematic endpoint $M_{T2}^{max}(M_C)$ will still vary 
with $\sqrt{\hat s}$ between a minimum value of 
\beq
M_{T2}^{max}(\sqrt{\hat s}=M_{A_1}+M_{A_2}) 
= \mu_{min} + \sqrt{\mu_{min}^2+M_C^2 }
\label{MT2maxmin}
\eeq
and a maximum value of 
\beq
M_{T2}^{max}(\sqrt{\hat s}\to\infty) 
= \mu_{max} + \sqrt{\mu_{max}^2+M_C^2 }.
\label{MT2maxmax}
\eeq

\begin{figure}[t] 
\begin{center} 
\includegraphics[width=9cm]{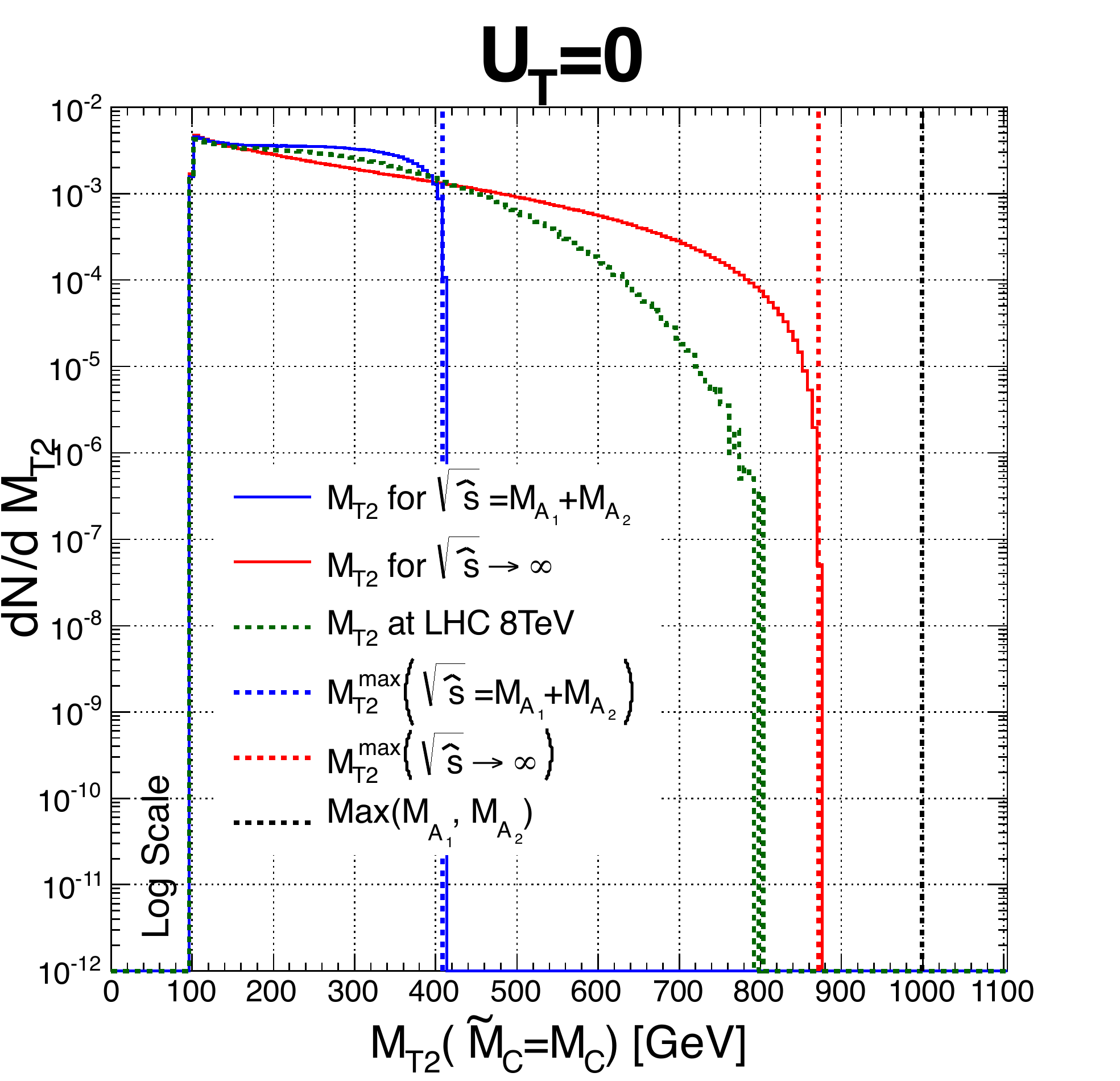} 
\end{center}
\caption{Unit-normalized $M_{T2}$ distributions for the asymmetric event topology of Fig.\,\ref{fig:DP}(a)
with different parent particles $A_1$ and $A_2$ and no upstream momentum ($U_T=0$). 
The particle mass spectrum is chosen as
$M_{A_1}=1000$ GeV, $M_{A_2}=200$ GeV and $M_{C}=100$ GeV. 
The $M_{T2}$ variable is computed with the correct value for the test mass $\tilde M_C=M_C=100$ GeV.
The blue histogram uses only events at threshold ($\sqrt{\hat s}=M_{A_1}+M_{A_2}$) and
its expected endpoint, marked by the vertical blue dashed line, is given by eq.\,(\ref{MT2maxblue}).
The red histogram shows the corresponding result in the infinite energy limit $\sqrt{\hat s}\to\infty$
(in practice, we take $\sqrt{\hat s}=100 M_{A_2}$) and the expected endpoint 
from eq.\,(\ref{MT2maxred}) is denoted by the vertical red dashed line.
The black vertical dashed line is the prediction from eq.\,(\ref{MT2maxDP}). 
The green dotted line is the result from {\tt Pythia} 6.4\,\cite{Sjostrand:2006za}
simulation at LHC8 for a realistic physics example corresponding to the event topology of Fig.\,\ref{fig:DP}(a):
associated squark-chargino ($\tilde q, \tilde \chi^\pm$) production, followed by
$\tilde q \to q+\tilde \chi^0$ and $\tilde \chi^\pm_1 \to \ell^\pm+\tilde \nu$, for the same 
mass spectrum, $(M_{\tilde q} , M_{\tilde\chi^\pm},M_{\tilde \nu}, M_{\tilde \chi^0} ) =(1000, 200, 100, 100 )$ GeV.
\label{fig:DPMT2} }
\end{figure}

This is illustrated in Fig.\,\ref{fig:DPMT2}, where we plot the $M_{T2}$ distributions for these two extreme cases:
$\sqrt{\hat s}=M_{A_1}+M_{A_2}$ (blue) and $\sqrt{\hat s}\to \infty$ (red).
We consider asymmetric events with different parents as in Fig.\,\ref{fig:DP}(a)
and choose the mass spectrum as follows $M_{A_1}=1000$ GeV,
$M_{A_2}=200$ GeV and $M_{C}=100$ GeV.
The blue and red vertical dashed lines mark the locations of the expected endpoints in eqs.\,(\ref{MT2maxmin}) and (\ref{MT2maxmax}), respectively. With the 
mass spectrum chosen for the figure, one gets
\bea
M_{T2}^{max}(\sqrt{\hat s}=M_{A_1}+M_{A_2}, \tilde M_C=M_C) &=&  409.8\ {\rm GeV}, 
\label{MT2maxblue}\\  
M_{T2}^{max}(\sqrt{\hat s}\to\infty, \tilde M_C=M_C)                      &=&  873.1\ {\rm GeV}.
\label{MT2maxred}
\eea
Fig.\,\ref{fig:DPMT2} demonstrates that in these two limiting cases, the endpoints of the 
$M_{T2}$ distributions agree perfectly with our expectations in eqs.\,(\ref{MT2maxblue}, \ref{MT2maxred}) and stay well below the conjecture of eq.\,(\ref{MT2maxDP}), which 
is indicated by the black vertical dashed line.

For intermediate, more realistic values of $\sqrt{\hat{s}}$, the upper endpoints of the
corresponding $M_{T2}$ distributions will populate  the region between those two extreme values,
but will certainly not exceed the theoretical maximum in eq.\,(\ref{MT2maxred}).
As an illustration, in Fig.\,\ref{fig:DPMT2} we also show results (the green dotted histogram)
from a realistic physics example simulated with {\tt Pythia} 6.4\,\cite{Sjostrand:2006za}. 
We considered associated squark-chargino production in supersymmetry, $pp\to \tilde q \tilde \chi^\pm$, followed by
$\tilde q \to q+\tilde \chi^0$ and $\tilde \chi^\pm_1 \to \ell^\pm+\tilde \nu$,
with the same mass spectrum as before,
$(M_{\tilde q} , M_{\tilde\chi^\pm},M_{\tilde \nu}, M_{\tilde \chi^0} ) =(1000, 200, 100, 100 )$ GeV.
Such events fall into the different parent category of Fig.\,\ref{fig:DP}(a).
With the available statistics, the $M_{T2}$ endpoint for the green dotted histogram
happens to be around 800 GeV, which, as expected, is in between 
eq.\,(\ref{MT2maxblue}) and eq.\,(\ref{MT2maxred}).
We see that the realistic $M_{T2}$ distribution indeed does not saturate the bound
of eq.\,(\ref{MT2maxDP}). Therefore, the correct interpretation of the $M_{T2}$ 
endpoint in the case of different parents should be made with the help of 
eqs.\,(\ref{eq:MT2maxDP}, \ref{eq:mueta}) instead.

\subsection{Events with upstream momentum ($U_T\ne0$)}
\label{sec:UT}

In the previous subsection~\ref{sec:UTzero}, we considered events with no upstream momentum ($U_T=0$),
where the parents $A_1$ and $A_2$ are produced back-to-back in the transverse plane.
In reality, however, the inclusive $(A_1, A_2)$ production is always associated with some 
amount of upstream momentum, either from initial state radiation or from decays of other, heavier particles. 
In this subsection we consider the effect of upstream momentum ($U_T\ne 0$) and show that 
our previous conclusions still hold.

It is well known that the kinematic endpoint $M_{T2}^{max}$ in general depends on the upstream 
momentum $U_T$, but this dependence is removed for a very special choice of the test masses, 
namely, when the test masses are equal to the true masses of the children particles, 
see eq.\,(\ref{eq:noUTdependence}).\footnote{The dependence on $U_T$ completely disappears for the 
case of the doubly projected $M_{T2\perp}$ variable introduced in \cite{Konar:2009wn}, 
which uses only the transverse momentum components orthogonal to $\vec{U}_T$.}
This property has been previously demonstrated only for the case of identical parent particles and now
we would like to test whether it also holds for the case of different parent particles in Fig.\,\ref{fig:DP}(a).

As we already explained, the case of different parents in Fig.\,\ref{fig:DP}(a) can be equivalently treated
as a case of identical parents, which decay asymmetrically as in Fig.\,\ref{fig:DP}(b).
In turn, this process can be described in terms of the asymmetric $M_{T2D}$ variable from 
eq.\,(\ref{eq:DLSPmt2}), whose endpoint $M_{T2D}^{max}(\tilde M_{\Psi_1}, \tilde M_{\Psi_2})$ will become 
independent of $U_T$ with the following choice of test masses
\beq
\tilde M_{\Psi_1} = \frac{M_{A_2}}{M_{A_1}}\, M_{C} , \qquad \tilde M_{\Psi_2}=M_{C}.
\label{eq:kinkposition}
\eeq
These values are in principle measurable experimentally, by studying the $U_T$ dependence of the 
$M_{T2D}^{max}$ endpoint as a function of the test masses, and finding the choice where
this dependence is minimized\,\cite{Konar:2009qr}.

Notice that if we try to use the symmetric version of the $M_{T2}(\tilde M_C)$ variable, where the 
input masses are equal ($\tilde M_{C_1}=\tilde M_{C_2}\equiv \tilde M_C$), we will get a less stringent bound. 
Even if the test mass is taken to be the true one ($\tilde M_C=M_C$), we still find a
chain of inequalities
\beq
 M_{T2}^{max} (M_C) < M_{T2D}^{max}\left( \frac{M_{A_2}}{M_{A_1}}\, M_{C} , M_{C}\right) \le M_{A_2} .
\label{eq:dpMT2}
\eeq

\begin{figure}[t!] 
\begin{center} 
\includegraphics[width=9cm]{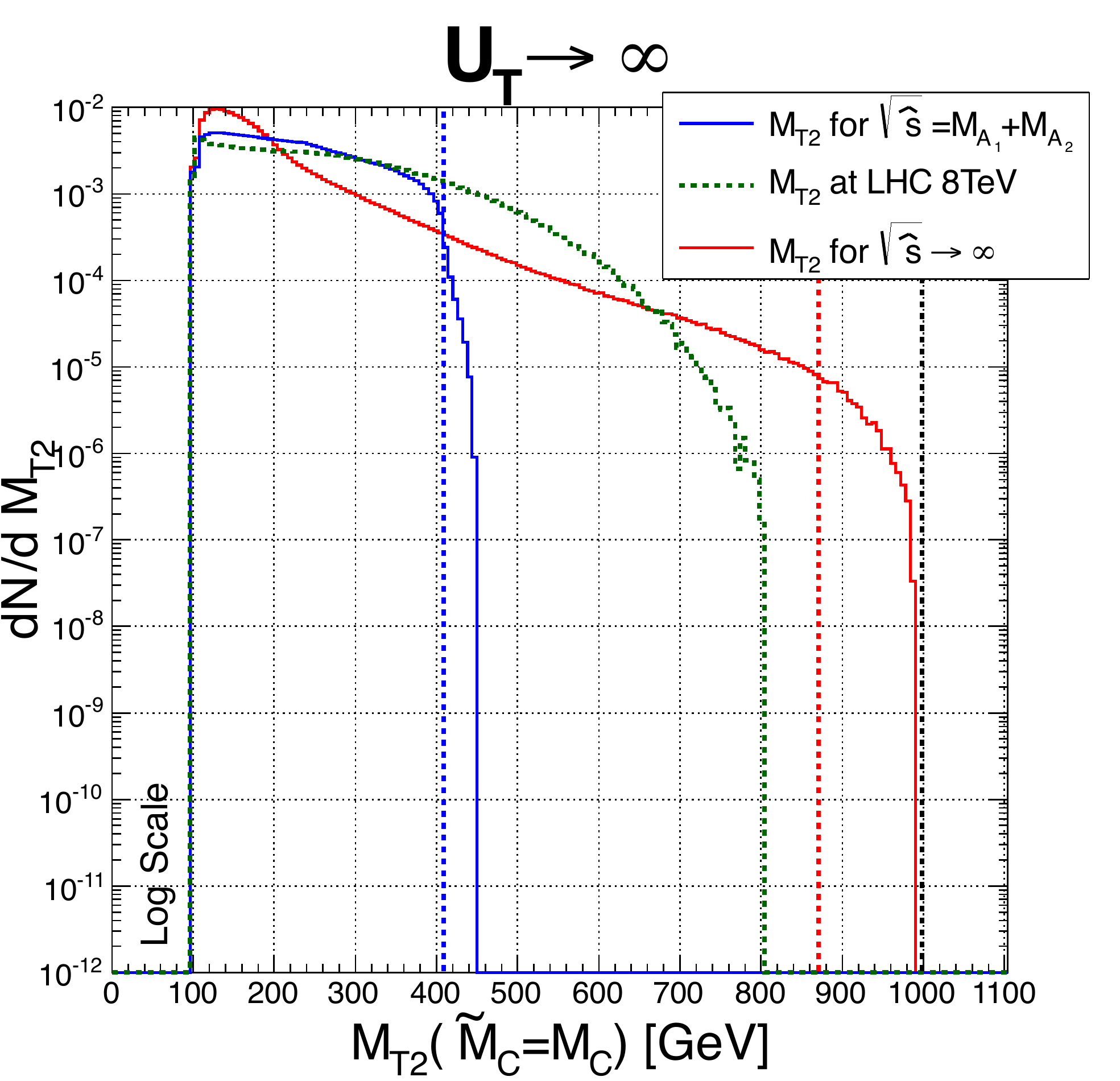}
\end{center}
\caption{The same as Fig.\,\ref{fig:DPMT2}, but for events with upstream momentum $U_T$. 
For the threshold limit ($\sqrt{\hat s}=M_{A_1}+M_{A_2}$) and
for the infinite energy limit ($\sqrt{\hat s}\to\infty$), large $U_T$ was put in by hand, while
for the realistic LHC 8 TeV simulation, $U_T$ was generated from initial state radiation in Pythia 6.4.
\label{fig:DPMT2UT} }
\end{figure}

Of course, when the test masses $\tilde M_{\Psi_1}$ and $\tilde M_{\Psi_2}$ are chosen away from the
special values in eq.\,(\ref{eq:kinkposition}), the endpoint $M_{T2D}^{max}$ as usual will be an increasing function of $U_T$. 
The same will also be true for its symmetric counterpart $M_{T2}^{max}$, whose $U_T$ dependence 
is illustrated in Fig.\,\ref{fig:DPMT2UT}, which is the analogue of Fig.\,\ref{fig:DPMT2} in the presence of
upstream momentum $U_T$. As before, the vertical blue and red dashed lines mark the locations
of the expected endpoints in eqs.\,(\ref{MT2maxblue}, \ref{MT2maxred}) in the absence of $U_T$.
The figure shows that, as expected, in the presence of upstream momentum, the endpoints are shifted higher,
and in the extreme case of infinite $\sqrt{s}$, the endpoint eventually reaches the naive 
expectation from eq.\,(\ref{MT2maxDP}):
\beq
\lim_{\sqrt{\hat s}, U_T \to \infty } M_{T2}^{max}(\tilde M_C=M_C) = M_{A_2}. 
\label{eq:DAKINK}
\eeq
Note that this limit is reached only in the unphysical case
when both $\sqrt{\hat s}$ and $U_T$ are sent to infinity.
In the realistic simulation of squark-chargino production (the green dotted line),
where $U_T$ is generated from initial state radiation, the $M_{T2}$ endpoint is 
similar to the one observed in Fig.\,\ref{fig:DPMT2} and again does not saturate the bound
eq.\,(\ref{eq:DAKINK}).

\subsection{Application to associated gluino-LSP production}

So far we have not at all discussed the event topologies in Figs.\,\ref{fig:diagrams}(k)
and \ref{fig:diagrams}(l), which can describe, e.g.~associated gluino-neutralino
production, where the gluino decay to the LSP gives 2 jets. Those topologies 
can be thought of as extreme examples of the 
``different parent" case just considered.  Thus,  $M_{T2}^{max}$ will again depend 
on the center of mass energy $\sqrt{\hat s}$ of the $AC_2$ system, which at hadron 
colliders will vary from one event to another. This will lead to a smearing of the endpoint 
$M_{T2}^{max}$. Thus we will consider the situation where the production energy $\sqrt{\hat s}$ of 
particles $A$ and $C_2$ is fixed, as at a future linear collider.
 
\underline{The event topology of Fig.\,\ref{fig:diagrams}(k).}
The $U_T$-invariant point of $M_{T2D}^{(max)}$ will appear at 
\beq
\left(\tilde M_{\Psi_1},\, \tilde M_{\Psi_2} \right) = \left(M_{\Psi_1}, \, M_{\Psi_2}\right), 
\eeq
where
\bea
M_{\Psi_1}&=&\frac{\sqrt{\hat s}}{2}\left\{1-\frac{2M_A}{\sqrt{\hat s}} \left(1-\frac{M_{B_1}^2}{M_A^2}\right) e^{\eta_s}\right\}^{1/2}, \\
M_{\Psi_2}&=&\frac{\sqrt{\hat s}}{2}\left\{1-\frac{2M_{B_1}}{\sqrt{\hat s}} \left(1-\frac{M_{C_1}^2}{M_{B_1}^2}\right) e^{\eta_s-\eta_b}\right\}^{1/2},\\
\eta_s&=&\cosh^{\minus 1}\left({\frac{\hat s+M_A^2-M_{C_2}^2}{2 \sqrt{\hat s}\, M_A}}\right), \label{eq:boosteta}\\
\eta_b&=&\cosh^{\minus 1}\left({\frac{M_A^2+M_{B_1}^2-M_{C_1}^2}{2 M_A\, M_{B_1}}}\right).
\eea
The corresponding value of $M_{T2D}^{(max)}$ will be
\beq
M_{T2D}^{(max)}\left(M_{\Psi_1}, M_{\Psi_2} \right) =\frac{\sqrt{\hat s}}{2}.
\eeq

\underline{The event topology of Fig.\,\ref{fig:diagrams}(l).}  With the effective topology 
technique, we find the $U_T$-invariant point at
\bea
\left(\tilde M_{\Psi_1},\, \tilde M_{\Psi_2} \right) &=& \left(M_\Psi, \, M_\Psi\right), \\
M_\Psi&=&\frac{\sqrt{\hat s}}{2}\left\{1-\frac{M_A}{\sqrt{\hat s}}\left(1-\frac{M_{C_1}^2}{M_A^2}\right)e^{\eta_s}\right\}^{1/2},  
\eea
with $\eta_s$ still given by eq.\,(\ref{eq:boosteta}). The endpoint is found at
\beq
M_{T2D}^{max}\left(M_\Psi, M_\Psi \right)= M_{T2}^{max}\left(M_\Psi\right)=\frac{\sqrt{\hat s}}{2}.
\eeq

\section{The shapes of $M_{T2_\perp}$ distributions}
\label{sec:shapes}

Until now we have been focusing on the measurable kinematic {\em endpoints} of 
different variables. At the same time, one could also attempt to study the {\em shapes}
of the corresponding differential distributions. Unfortunately, (to the best of our knowledge)
analytical formulas for the shapes of the $M_{T2}$ and $M_{T2D}$ distributions are absent.
Their derivation would be rather complicated, because the shapes are affected by several
factors: the production energy $\sqrt{\hat s}$, spin correlations, upstream 
momentum $U_T$, etc. In order to remove the $U_T$ effect, Ref.\,\cite{Konar:2009wn}
introduced a 1D-projection of $M_{T2}$, called  $M_{T2_\perp}$.  
$M_{T2_\perp}$ is calculated the same way as $M_{T2}$, except that it
uses the projections of the transverse momenta on the dimension which is orthogonal 
to the $\vec{U}_T$ direction\,\cite{Matchev:2009ad}. It turned out that this 
doubly projected variable is also independent of $\sqrt{\hat s}$ and avoids large
spin correlation effects\,\cite{Edelhauser:2012xb}. 
In this section we provide analytical formulas for the shapes of the $M_{T2_\perp}$
distributions for various event topologies from Fig.\,\ref{fig:diagrams}. 
We will also discuss the endpoint behavior of these distributions, extending
the technique proposed in\,\cite{Giudice:2011ib} to count the number of invisible 
particles $N_{inv}$.

With two massless visible particles,   $M_{T2_\perp}{\scriptstyle\left(\tilde M_{\Psi}\right)}$  is related to  
$M_{T2_\perp}{\left(0\right)}$ as follows
 \beq
x  \equiv  M_{T2_\perp}{\textstyle\left(0\right)} =
\frac{M_{T2_\perp}{\scriptstyle \left(\tilde M_\Psi\right)}^2-\tilde M_\Psi^2}{M_{T2_\perp}{\scriptstyle \left(\tilde M_{\Psi}\right)}}.
\eeq
The corresponding distributions are related as
\beq
\frac{\ud N}{\ud M_{T2_\perp}{\scriptstyle \left(\tilde M_\Psi\right)}} = 
\left(\frac{M_{T2_\perp}{\scriptstyle \left(\tilde M_\Psi\right)}^2+\tilde M_\Psi^2}{M_{T2_\perp}{\scriptstyle \left(\tilde M_{\Psi}\right)}^2}\right)
 \cdot \frac{\ud N}{\ud x}
\label{eq:mt2perpMT2}
\eeq 
Thus we only need to describe the shape of $x$, since  the shape of
$M_{T2_\perp}{\scriptstyle \left(\tilde M_\Psi\right)}$ will then be easily 
obtained from eq.\,(\ref{eq:mt2perpMT2}).
In the following equations, we used the same notation for $\mu$ and $\eta_i$
as in eqs.\,(\ref{mudef}, \ref{eq:eta}), respectively.  
We also introduce individual $\mu_i$ defined as
 \beq
 \mu_i = \frac{M_A}{2}\left(1-\frac{M_{\Psi_i}^2}{M_A^2}\right), 
 \eeq
so that $\mu$ is the geometric mean of $\mu_1$ and $\mu_2$, as in eq.\,(\ref{mudef}),
 \beq
\mu = \sqrt{\mu_1 \mu_2}. 
 \eeq
 In the following we list our results for $dN/dx$.

\underline{The event topologies of Figs.\,\ref{fig:diagrams}(a), \ref{fig:diagrams}(b) and \ref{fig:diagrams}(e).} 
The answer is very simple\,\cite{Konar:2009wn}
 \beq
 \frac{\ud N}{\ud x} \propto x\,  \log\left(\frac{2 \mu }{x }\right). 
\label{dNdxabe}
 \eeq

\underline{The event topologies of Figs.\,\ref{fig:diagrams}(c) and \ref{fig:diagrams}(f).} We find
\beq
\frac{\ud N}{\ud x}\propto x \int_\frac{x}{2 \mu}^{1} \frac{\ud p}{p} \, J_{\textrm{on}}^{(1)}\left(p\right) ,
\label{dNdxcf_formula}
\eeq
where
\beq
J_{\textrm{on}}^{(i)}\left(p\right)\equiv  \eta_i-\Theta\left(p-e^{-\eta_i}\right) \ln\left(p\,e^{\eta_i}\right).
\label{eq:onJ}
\eeq
Here $\Theta\left(x\right)$ is a unit step function and $J_{\textrm{on}}^{(i)}\left(p\right)$ is a phase 
space weight  from the cascade decay chain.
After integrating out eq.\,(\ref{dNdxcf_formula}), we get 
 \beq
  \frac{\ud N}{\ud x} \propto x\cdot
  \begin{cases}
 2\eta_1  \ln\left(\frac{2  \mu}{x\, e^{\frac{\eta_1}{2}}}\right)   \textrm{ if  } 0\le x<2 \mu\, e^{-\eta_1}, \\[5mm]
\left[ \ln\left(\frac{2 \mu }{x }\right)\right]^2  \textrm{ if } 2 \mu\, e^{-\eta_1} \le x \le 2 \mu .  
 \end{cases}
 \label{dNdxcf}
  \eeq

\underline{The event topologies of Figs.\,\ref{fig:diagrams}(d) and \ref{fig:diagrams}(g).} 
We find
 \bea 
  \frac{d N}{d x}\propto && x \int_\frac{x}{2 \mu}^{1} \frac{\ud q}{q} \, J_{\textrm{off}}^{(1)}\left(q\right) ,
  \\
    J_{\textrm{off}}^{(i)}\left(q\right)\equiv &&\int_{ {  1}}^{\frac{M_A^2}{M_{\Psi_i}^2} (1-q^2)+q^2}
\frac{ \ud s}{s} \sqrt{(s-1)
\left(s-\frac{ 
{\textstyle |M_{C_i}-M_{\chi_i}|^2}}{M_{\Psi_i}^2}\right)}.
\label{eq:offJ}
\eea
The function $ J_{\textrm{off}}^{(i)}\left(q\right)$ is a three-body phase space weight. Integrating eq.\,(\ref{eq:offJ}) above, we get
\bea
J_{\textrm{off}}^{(i)}\left(q\right)  &=& \frac{M_A^2-M_{\Psi_i}^2}{M_{\Delta_i}^2} \sqrt{R_i - q^2}\sqrt{1-q^2}
-\frac{M_{\Psi_i}}{M_{\Delta_i}} \log\left(\frac{M_{\Psi_i} \sqrt{R_i-q^2} - M_{\Delta_i} \sqrt{1-q^2}}
 {M_{\Psi_i} \sqrt{R_i-q^2} +M_{\Delta_i} \sqrt{1-q^2}}\right)
\nonumber \\
&-&\left(\frac{M_{\Psi_i}^2+M_{\Delta_i}^2}{M_{\Delta_i}^2} \right) 
\Bigg\{\frac{1}{2}\log\left(\frac{M_A^2-M_{\Psi_i}^2}{M_{\Psi_i}^2-M_{\Delta_i}^2}\right)   
+\log\left(\sqrt{R_i-q^2}+\sqrt{1-q^2}\right)\Bigg\}, ~~~~~~
 \eea
with
\beq
M_{\Delta_i} = | M_{C_i}-M_{\chi_i}| , \quad R_i =  \frac{M_A^2- M_{\Delta_i}^2}{M_A^2- M_{\Psi_i}^2}.
\eeq
When $\chi_i$ is massless, 
\beq
J_{\textrm{off}}^{(i)}\left(q\right) = \left(\frac{M_A^2-M_{C_i}^2}{M_{C_i}^2} \right)(1-q^2)-\ln\left[\frac{M_A^2}{M_{C_i}^2}(1-q^2)+q^2\right],
\eeq
and the corresponding distribution becomes
 \bea
 \frac{\ud N}{\ud x} &\propto&  x   \Bigg[ 
  \frac{M_A^2-M_{C_1}^2}{M_{C_1}^2} 
  \left\{ -1+\frac{x^2}{4 \mu^2}
  -2 \ln\left(\frac{x}{2 \mu}\right)\right\}}  {\textstyle - 4 \ln\left(\frac{M_{C_1}}{M_A}\right) \ln\left(\frac{x}{2 \mu}\right) \nonumber \\
&&\quad +\textrm{Li}_2\left(\frac{2 \mu_1}{M_A}\right)   -\textrm{Li}_2\left(\frac{x^2}{2 \mu_2 M_A}\right) \Bigg],
  \eea
 where $\textrm{Li}_2\left(x\right)$ is Spence's function, defined as
 \beq
 \textrm{Li}_2\left(x\right)=-\int_0^x \ud z \frac{\ln\left(|1-z|\right)}{z}.
 \eeq

\underline{The event topology of Fig.\,\ref{fig:diagrams}(h).}
Without loss of generality, $\eta_2 \le \eta_1$ and we have
  \beq
\frac{\ud N}{\ud x} \propto x \cdot 
   \begin{cases}
 \ln\left(\frac{2 \mu}{x e^{\frac{\eta_1+\eta_2}{2}}}\right) \textrm{ if } 0\le x < x_0, \\[3mm]
 \ln\left(\frac{2 \mu}{x e^{\frac{\eta_1+\eta_2}{2}}}\right)
  -\frac{1}{6 \eta_1 \eta_2} \left[\ln\left(\frac{2 \mu}{x e^{\eta_1+\eta_2}}\right)\right]^3  \textrm{ if } x_0 \le x < x_1,\\[4mm]
 \frac{\eta_2}{6\eta_1} 
 \left(\eta_2-3 \ln\left(\frac{2\mu}{x}\right)+\frac{3}{\eta_2}  \left[\ln\left(\frac{2 \mu}{x}\right)\right]^2\right)  
 \textrm{ if } x_1 \le x < x_2, \\[3mm]
 \frac{1}{6 \eta_1 \eta_2}\left[\ln\left(\frac{2 \mu}{x}\right)\right]^3 \textrm{ if } x_2\le x \le x_3 ,
 \end{cases}
  \label{dNdxh}
 \eeq    
 where
 \bea
 x_0=2 \mu\, e^{-\left(\eta_1+\eta_2\right)}, &&\quad x_1= 2 \mu\, e^{-\eta_1},\\
 x_2=  2 \mu\, e^{-\eta_2},&&\quad x_3=2 \mu.
 \eea

\underline{The event topology of Fig.\,\ref{fig:diagrams}(i).} We find
 \beq 
 \frac{\ud N}{\ud x}  \propto  x  \cdot
 \begin{cases} 
\int_{\frac{x}{2 \mu}}^{\frac{x e^{\eta_1}}{2 \mu}} 
 \frac{\ud q}{q}  \ln\left(\frac{2 \mu \, q}{x}\right)  J_{\textrm{off}}^{(2)}\left(q\right)  
  +    \int_{\frac{x e^{\eta_1}}{2 \mu}}^{1} 
 \frac{\ud q}{q} (\eta_1) \,J_{\textrm{off}}^{(2)}\left(q\right) \, ,  \textrm{  if } 0 \le x <x_1,\\[6mm]
   \int_{\frac{x}{2 \mu} }^{1} 
 \frac{\ud q}{q}  \ln\left(\frac{ 2 \mu \, q}{x}\right)  J_{\textrm{off}}^{(2)}\left(q\right) \, , \textrm{ if } x_1 \le x \le x_3.
 \end{cases}
 \eeq
 
 \underline{The event topology of Fig.\,\ref{fig:diagrams}(j).} We leave it in integral form
 \beq
 \frac{\ud N}{\ud x}  \propto  x \int_{\frac{x}{2 \mu}}^{1} \frac{\ud q}{q} 
 J_{\textrm{off}}^{(1)}\left(q\right) J_{\textrm{off}}^{(2)}\left(\frac{x}{2 \mu \,q}\right) .
 \eeq
 
\begin{figure}[t] 
\begin{center}
\includegraphics[width=9cm]{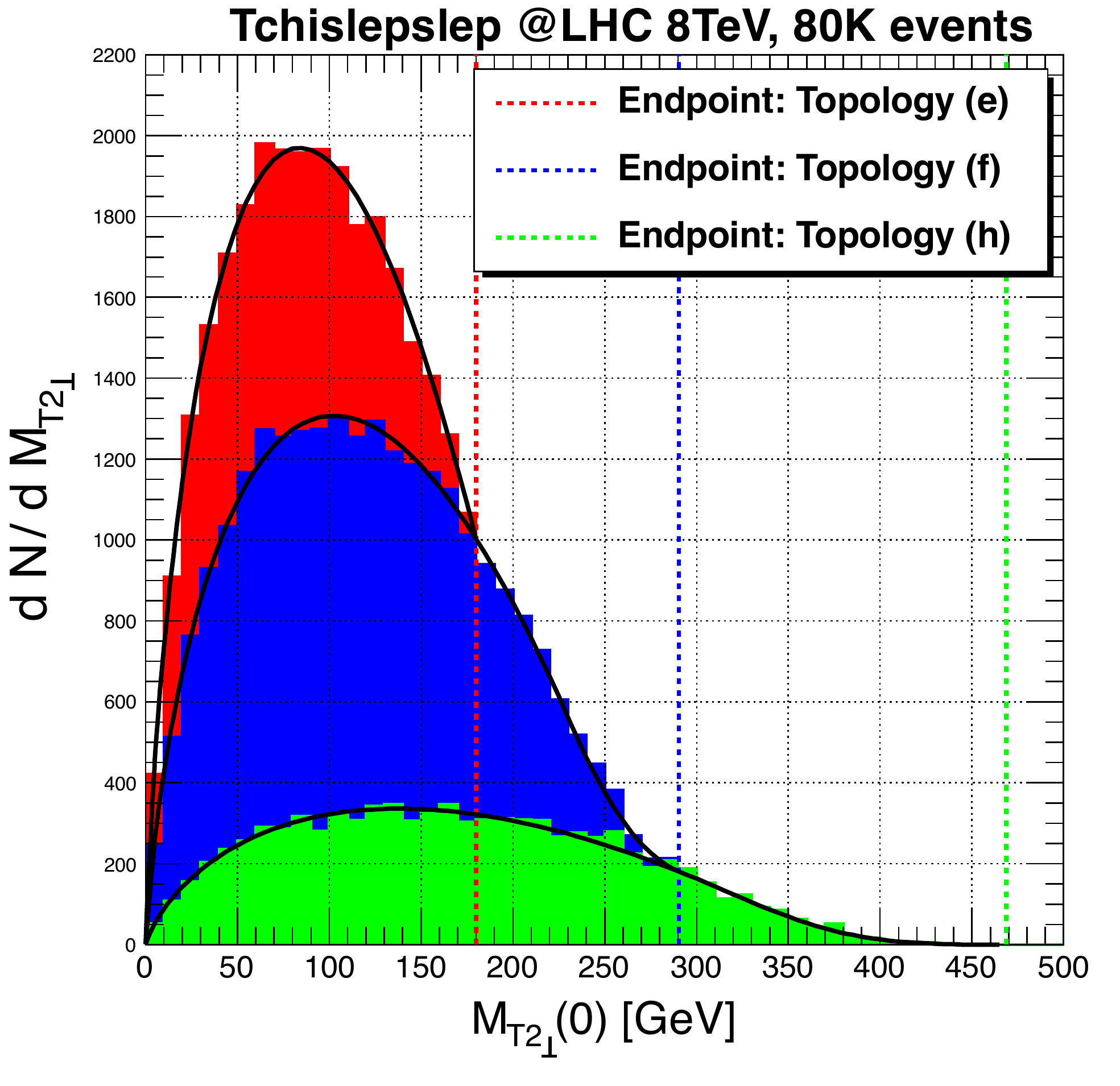}
\end{center}
\caption{The same as Fig.\,\ref{fig:LM7}, but for the doubly projected $M_{T2}$ variable $M_{T2_\perp}$.
The solid lines show the corresponding theoretical predictions following from eqs.\,(\ref{dNdxabe}, \ref{dNdxcf}, \ref{dNdxh}).
\label{fig:LM7D} }
\end{figure}
 
For illustration, in Fig.\,\ref{fig:LM7D} we show the corresponding $M_{T2_\perp}$ distributions 
for the {\tt Tchislepslep} SMS model considered in Fig.\,\ref{fig:LM7}. In addition to the histograms which
were obtained from numerical simulations, we also show the corresponding theoretical predictions 
following from eqs.\,(\ref{dNdxabe}, \ref{dNdxcf}, \ref{dNdxh}). We see that the analytical results
match very well the simulated $M_{T2_\perp}$ differential distributions.

As detailed in \cite{Giudice:2011ib}, it should, in principle, be possible to distinguish between topologies
with different numbers of invisible particles $N_{inv}$ simply by fitting the endpoint fall-off of kinematic distributions; this was shown to have a near-universal dependence on $N_{inv}$ for judiciously chosen variables\,\cite{Giudice:2011ib}.  The near-endpoint behavior of the doubly-projected variable $M_{T2_\perp}$ for the various topologies in Fig.\,\ref{fig:diagrams} is detailed in Table\,\ref{tab:endpoint} as a function of$\epsilon$, the distance away from the endpoint. Note firstly that the fall-off for the doubly-projected variable $M_{T2_\perp}$ is faster than that for the usual stransverse mass $M_{T2}$.  This is due to an additional dimension of the full phase space being `projected out', and making it more difficult to distinguish
between different numbers of invisibles $N_{inv}$ for large $N_{inv}$ using the endpoint behavior alone.  Secondly, although the near-endpoint behavior has a universal dependence on $N_{inv}$ for massless invisible particles\,\cite{Giudice:2011ib}, this is not true when any invisible particles obtain a non-negligible mass.  In the latter case we see that the fall-off is always slower for true cascade decays (with on-shell resonances), since the presence of an on-shell intermediate particle effectively reduces the dimension of the full phase space, hence fewer dimensions are projected out.  In fact, for cascades, the near-endpoint behavior is {\it entirely independent} of the invisible particle masses. Finally, note that there are discrete ambiguities between the endpoint fall-off of cascades, and decays with off-shell intermediates, for different $N_{inv}$.  In these particular cases, then, one would need to fit using the full shape formula, given above.
\begin{table}
\centering
\begin{tabular}{|c|c|c||c|c|}
\hline
\multirow{2}{*}{$N_{inv}$}&\multicolumn{2}{c||}{On-shell topologies}&
\multicolumn{2}{c|}{Off-shell topologies}\\
&topology & near-endpoint behavior & topology & near-endpoint behavior \\
\hline
\hline
2  & (a)      & $\epsilon$   &                &                   \\
\hline
3  & (c),(f)  & $\epsilon^2$ & (d),(g)        & $\epsilon^3$ \\
\hline
\multirow{2}{*}{4}  & (h)      & $\epsilon^3$ & (i)            & $\epsilon^4$ \\
 &          &              & (j)            & $\epsilon^5$  \\
\hline
\end{tabular}
\caption{\label{tab:endpoint}Near-endpoint behavior of the doubly-projected stransverse mass $M_{T2\perp}$ distribution, for all topologies in Fig.\,\ref{fig:diagrams}, as a function of $\epsilon$, the distance away from the endpoint.  It is assumed that one invisible particle on each leg of the decay is massive (corresponding to the LSP), while any others are massless (corresponding to additional neutrinos emitted in the decay).  The fall-off for alternative mass spectra can be trivially obtained by taking the relevant limit in the shape expressions given above.}
\end{table}

\section{Conclusions and summary}
\label{sec:conclusions}

Our work in this paper removes some of the restrictions which so far have prevented
the more widespread usage of the $M_{T2}$ variable. We demonstrated how the 
$M_{T2}$ variable (and its variants) can be usefully applied in more general situations, e.g.:
\begin{itemize}
\item \underline{Decay chains with multiple invisible particles.} In Sections~\ref{sec:3plus1} 
and \ref{sec:effective} we considered cases where the new physics decay chain gives rise to
{\em several} invisible particles. Previous $M_{T2}$ studies have typically assumed that
there is only one invisible particle in the decay chain (the dark matter WIMP), which 
appears at the end of the decay chain. At the same time, there are many scenarios in which
{\em additional} invisible particles can be present. The most popular example of this sort 
are chargino decays in supersymmetry, which yield SM neutrinos {\em in addition} to the
invisible LSP. We proposed two methods for dealing with the problem of additional invisibles:
by introducing topology-dependent new variables (in Section~\ref{sec:3plus1})
and by reinterpreting the measured kinematic endpoints of the conventional 
$M_{T2}$ distributions (in Section~\ref{sec:effective}).
\item \underline{Events with different invisible child particles.} In Section~\ref{sec:DLSP}
we pointed out that the reinterpretation method carries over to the case where 
the invisible child particles at the end of the decay chains are different. The key idea 
is to use the asymmetric $M_{T2D}$ variable introduced in \cite{Konar:2009qr}.
\item \underline{Events with different parent particles.} Much of the previous literature 
on $M_{T2}$ dealt only with events in which the two parent particles initiating
the decay chains are identical. As for events with different parent particles, it was 
thought that the $M_{T2}$ endpoint in that case reveals the mass of the {\em heavier} parent.
In Section~\ref{sec:DP} we showed that this conjecture is false, and we 
gave the correct interpretation of the $M_{T2}$ endpoint for the case of different parent particles.
\end{itemize}

Apart from a good theoretical understanding of the measured $M_{T2}$ endpoint in terms of the underlying
mass spectrum in all those different situations, it is also important to have a good knowledge of the 
shapes of the respective differential $M_{T2}$ distributions. In Section~\ref{sec:shapes} 
we considered the doubly projected variable $M_{T2_\perp}$ proposed in \cite{Konar:2009wn}
and derived the corresponding shapes for a number of different cases shown in Fig.\,\ref{fig:diagrams}(a-j).
Our formulas can be used for improving the precision of mass measurements based on 
$M_{T2_\perp}$ kinematic endpoints\,\cite{CMStop}. 
Furthermore, by comparing the different shape predictions to the data, one could also, 
in principle, deduce the correct event topology and/or the number of invisible particles 
in the event, although in practice this may be unfeasible due to limited statistics near the endpoint.

\acknowledgments
We thank W.~Cho, K.~Kong, F.~Moortgat and L.~Pape for stimulating discussions.
R.~Mahbubani thanks the CERN TH group for its warm hospitality.
K.~Matchev is supported in part by a U.S.~Department of Energy grant DE-FG02-97ER41029.
M.~Park is supported by the CERN-Korea fellowship through National Research Foundation of Korea.

\end{document}